\documentclass[camera,hyphens]{jpaper}
\usepackage{graphicx}
\usepackage{booktabs}
\usepackage{verbatim}
\usepackage{bm}
\usepackage{listings}
\usepackage{caption}
\usepackage{balance}
\usepackage{threeparttable,booktabs}
\usepackage[algoruled,resetcount,linesnumbered ]{algorithm2e}

\usepackage{float}
\usepackage{amsmath}
\usepackage{tikz}
\usetikzlibrary{automata,arrows,topaths, shapes}
\usepackage{siunitx}
\usepackage{hyperref}
\usepackage{subcaption}
\usepackage{parcolumns}
\usepackage{soul}
\usepackage{eso-pic,lipsum}
\hypersetup{
    colorlinks=false,
    pdfborder={0 0 0},
}

\usepackage{ragged2e}   
\usepackage{glossaries}
\usepackage{booktabs}

\usepackage{multirow}

	\makeatletter
	\g@addto@macro{\normalsize}{%
	  \setlength{\abovedisplayskip}{2pt plus 1pt minus 1pt}
	  \setlength{\belowdisplayskip}{2pt plus 1pt minus 1pt}
	  \setlength{\abovedisplayshortskip}{0pt}
	  \setlength{\belowdisplayshortskip}{0pt}
	  \setlength{\intextsep}{2pt plus 1pt minus 1pt}
	  \setlength{\textfloatsep}{4pt plus 1pt minus 1pt}
	  \setlength{\skip\footins}{5pt plus 1pt minus 1pt}}
	\makeatother
	\usepackage{titlesec}

\makeatletter
\def\endthebibliography{%
  \def\@noitemerr{\@latex@warning{Empty `thebibliography' environment}}%
  \endlist
}
\usepackage{setspace}

\usepackage{cite}
\usepackage{amsmath,amssymb,amsfonts}
\usepackage{algorithmic}
\usepackage{graphicx}
\usepackage{textcomp}
\usepackage{xcolor}
\usepackage{sidecap}

\usepackage{tikz}

\usepackage{enumitem}

 \usepackage{fancyhdr}

\def\BibTeX{{\rm B\kern-.05em{\sc i\kern-.025em b}\kern-.08em
    T\kern-.1667em\lower.7ex\hbox{E}\kern-.125emX}}

\usepackage[hang,flushmargin]{footmisc}
\newcommand*\circled[1]{\tikz[baseline=(char.base)]{
            \node[shape=circle,draw,inner sep=0pt,fill=black, text=white] (char) {#1};}}

\definecolor{dred}{rgb}{0.75, 0.00, 0.00}
\definecolor{dgreen}{rgb}{0.00, 0.5, 0.00}
\definecolor{ddgreen}{rgb}{0.00, 0.50, 0.00}
\definecolor{dpink}{rgb}{0.75, 0.0, 0.75}
\definecolor{dblack}{rgb}{0.00, 0.00, 0.00}
\definecolor{dblue}{rgb}{0.00, 0.00, 0.75}
\definecolor{gyell}{rgb}{0.5, 0.5, 0.0}
\definecolor{dbleudefrance}{rgb}{0.19, 0.55, 0.91}
\definecolor{darkgoldenrod}{rgb}{0.72, 0.53, 0.04}
\definecolor{magenta}{rgb}{160,0,255}

\newcommand{\juan}[1]{{\color{black}#1}}
\newcommand{\juang}[1]{{\color{black}#1}}
\newcommand{\gagan}[1]{{\color{black}#1}}
\newcommand{\gs}[1]{{\color{black}#1}}
\newcommand{\gss}[1]{{\color{black}#1}}
\newcommand{\om}[1]{{\color{black}#1}}
\newcommand{\omm}[1]{{\color{black}#1}}
\newcommand{\ommm}[1]{{\color{black}#1}}
\newcommand{\omu}[1]{{\color{black}#1}}
\newcommand{\gaganT}[1]{{\color{black}#1}}

\renewcommand{\baselinestretch}{1}
\newcommand{\head}[1]{{\noindent\textbf{#1.}\xspace}} 
\newcommand{\namePaper}{LEAPER\xspace} 
\newcommand{\etal}{\textit{et al.}}
\begin{document}

\title{\namePaper: {Modeling \gagan{Cloud} FPGA-based} Systems\\via Transfer Learning}

\title{\namePaper: Fast and Accurate FPGA-based System\\ Performance Prediction via Transfer Learning}


\author{Gagandeep Singh$^{a}$ \hspace{0.5cm}  Dionysios Diamantopoulos$^b$ \hspace{0.5cm}  Juan G{\'o}mez-Luna$^a$ \\  Sander Stuijk$^c$ \hspace{1cm}  Henk Corporaal$^c$ \hspace{1cm}  Onur Mutlu$^a$ \\
\vspace{-0.4cm} \normalsize $^a$ETH Z{\"u}rich    \hspace{1cm}  $^b$IBM Research Europe, Zurich \hspace{1cm} $^c$Eindhoven University of Technology \vspace{0.4cm}}

\maketitle
\thispagestyle{plain}
\pagestyle{plain}
\begin{abstract}
Machine learning 
 \omm{has} recently gained traction as a way to overcome the slow \om{accelerator generation and implementation process on an FPGA}. 
\juang{It can be used to build} \om{performance and resource \ommm{usage}} models that \om{enable fast early-stage design space exploration}. 
However, these models suffer from \omm{three} main limitations. 
\juang{First,} training requires large amounts of data (features extracted from design synthesis and implementation tools), which is cost-inefficient because of the time-consuming \om{accelerator} design \om{and implementation process}. 
\juang{Second,} a model trained for a specific environment cannot predict \om{performance} 
\juang{or resource \ommm{usage}} for a new, unknown environment. In a cloud system, 
\om{renting a platform for} data collection \om{to build an} ML model can significantly increase the total-cost-ownership (TCO) of a system. \omm{Third, ML-based models trained using a limited number of samples are prone to overfitting.} To overcome these limitations, we propose \namePaper, a \textit{transfer learning}-based approach for \juang{prediction of performance and resource \ommm{usage} in} FPGA-based systems. 
\juang{The key idea of \namePaper is to transfer} 
an ML-based performance and resource \ommm{usage} model \om{trained for a low-end edge environment} to a new, high-end cloud environment to provide fast and accurate predictions \om{for 
accelerator implementation}. 
Experimental results show that \om{\namePaper} 
\juang{(1) provides}, on average \juang{across six workloads and five FPGAs}, 85\% accuracy {when we use our transferred model for prediction in a cloud environment} with \textit{5-shot learning} and \juang{(2)} reduces {design-space exploration} time \om{for 
accelerator implementation on an FPGA} by 10$\times$, from days to {only} a few hours.

\end{abstract}


\vspace{-0.08cm}
\section{Introduction}

The need for energy efficiency \om{and flexible acceleration of workloads} has boosted the widespread adoption of field-programmable gate arrays (FPGAs)~\cite{singh2021fpga,van2019coherently,singh2019narmada,singh2019low} \om{in both edge and cloud computing}. Past works~\cite{singh2021fpga,van2019coherently,singh2019narmada,singh2019low,jiang2020boyi,cali2022segram, dai2017foregraph,alser2017,alser2019sneakysnake,diamantopoulos2018ectalk,kara2018columnml,singh2022accelerating} show that FPGAs can be employed {effectively} \om{to accelerate} a wide range of applications, {including graph processing, databases, neural networks, 
weather forecasting, \juang{and genome analysis}}. 

An FPGA is highly configurable as its circuitry can be tailored to perform any task. \om{However, FPGA developers face two main issues while \juang{designing} an accelerator.} 
\om{First, the} large configuration space of \om{an} FPGA and {the} complex interactions among \om{its} configuration options \om{cause} many 
developers to explore optimization \om{techniques} in an ad-hoc manner~\cite{diamantopoulos2020agile,singh2021modeling}. 
\om{Second, FPGA programming leads to low productivity because of the time-consuming accelerator design and implementation} process~\cite{o2018predictive}. 
Therefore, a common challenge that past works {have faced} is how to evaluate the performance \juang{(and resource \omm{usage})} of an accelerator implementation in a reasonable amount of time~\cite{o2018hlspredict}. 
To overcome this problem, researchers have {recently} employed machine learning (ML)-based models~\cite{makrani2019pyramid,ferianc2020improving,ustun2020accurate,o2018hlspredict,xppe_asp_dac,dai2018fast,zhao2019machine,yanghua2016improving,wang2020machine,mahapatra2014machine,singh2019napel} to \om{
\juang{predict the performance and resource \ommm{usage} of a given} accelerator implementation quickly. 
However, these ML-based models have three fundamental issues that 
\juang{reduce their usability.}}

First, \om{these ML-based} 
\juang{predictors} are trained for specific workloads, fixed hardware, and/or a set of inputs. {Therefore,} we cannot reuse these models for 
\juang{a} \om{previously \textit{unseen}} workload or \om{a different FPGA platform} because the trained model does not have a notion of the new, {unknown} environment.\footnote{We consider a new workload or a \omm{new} FPGA platform as a \omm{new} environment.} 
Therefore, traditional ML-based models have limited \textit{reusability}.

Second, \om{ML-based models} require a 
\omu{large} {number} of samples to construct a useful \om{performance} predict\om{or}. Collecting such a large number of samples is often \om{very} time-consuming due to the very long \om{accelerator} implementation cycle \om{on an FPGA, especially in a cloud computing environment \omu{where} data collection could be costly.} 

Third, \om{ML-based} models trained using a limited \juang{number of} samples are prone to serious \emph{overfitting} problems \juan{(i.e., model matches to the training data \om{too closely)}}~\cite{dai2007boosting}, limiting model generalization. 

\textbf{Our goal} is \juang{to overcome these three issues of ML-based models for prediction of FPGA performance and resource \ommm{usage}.} 
\juang{To this end, we present \namePaper. Our key idea is} 
to leverage an ML-based performance and resource \ommm{usage} model trained for a low-end \om{edge environment} to predict performance and resource \ommm{usage} of an accelerator implementation \om{for a new, high-end cloud environment.} 

\namePaper\footnote{We call our mechanism \namePaper~because it allows us to hop or ``leap'' between machine learning models.} \juang{employs} a transfer learning-based approach 
\juang{(also called \textit{few-shot learning}~\cite{fsl}). 
This technique is based on the idea that algorithms, similar to humans, can learn from past experiences and \emph{transfer} knowledge to the resolution of previously-unknown tasks.} 
\juang{Concretely, \namePaper transfers an ML-based model trained on an edge FPGA-based system to a new, high-end cloud FPGA-based system.} 
\gs{Using an edge FPGA-based system for 
\juang{training the ML-based model provides three major benefits over using a high-end FPGA.} 
First, since 
edge devices are small, FPGA bitstream generation is faster compared to generating bitstream for a high-end FPGA. 
Second, edge FPGAs are cheaper and more affordable. 
Third, a high-end FPGA often requires integration with a server-grade host CPU, which can be 
\juang{costly or not possible} for many users. 
Therefore, using low-cost and broadly available edge systems for \juang{training} data collection can} 
\juang{greatly facilitate the generation of ML-based predictors for performance and resource \ommm{usage} of FPGA-based systems.} 

\juang{\namePaper consists of three main steps.} 
\juan{First,} \namePaper uses \emph{design of experiments} (\emph{DoE})~\cite{montgomery2017design}, \juang{a technique} to extract representative \juang{training} data 
\juang{from} a small number of experimental runs. 
\juan{Second,} 
\juang{\namePaper trains} an ML-based model (\textit{base model}) to predict performance or resource \ommm{usage} for an accelerator implementation on a low-end edge environment. 
\juan{Third,} 
\juang{\namePaper transfers the} 
trained base model 
to 
a new, high-end \emph{cloud} environment (a new FPGA or a new application) with only a few \juang{new training samples (between 5 to 10 samples) from the new environment}. 

\begin{figure}[h]
  \centering
  \includegraphics[width=0.9\linewidth,trim={0.2cm 0.45cm 0.4cm 0.3cm},clip]{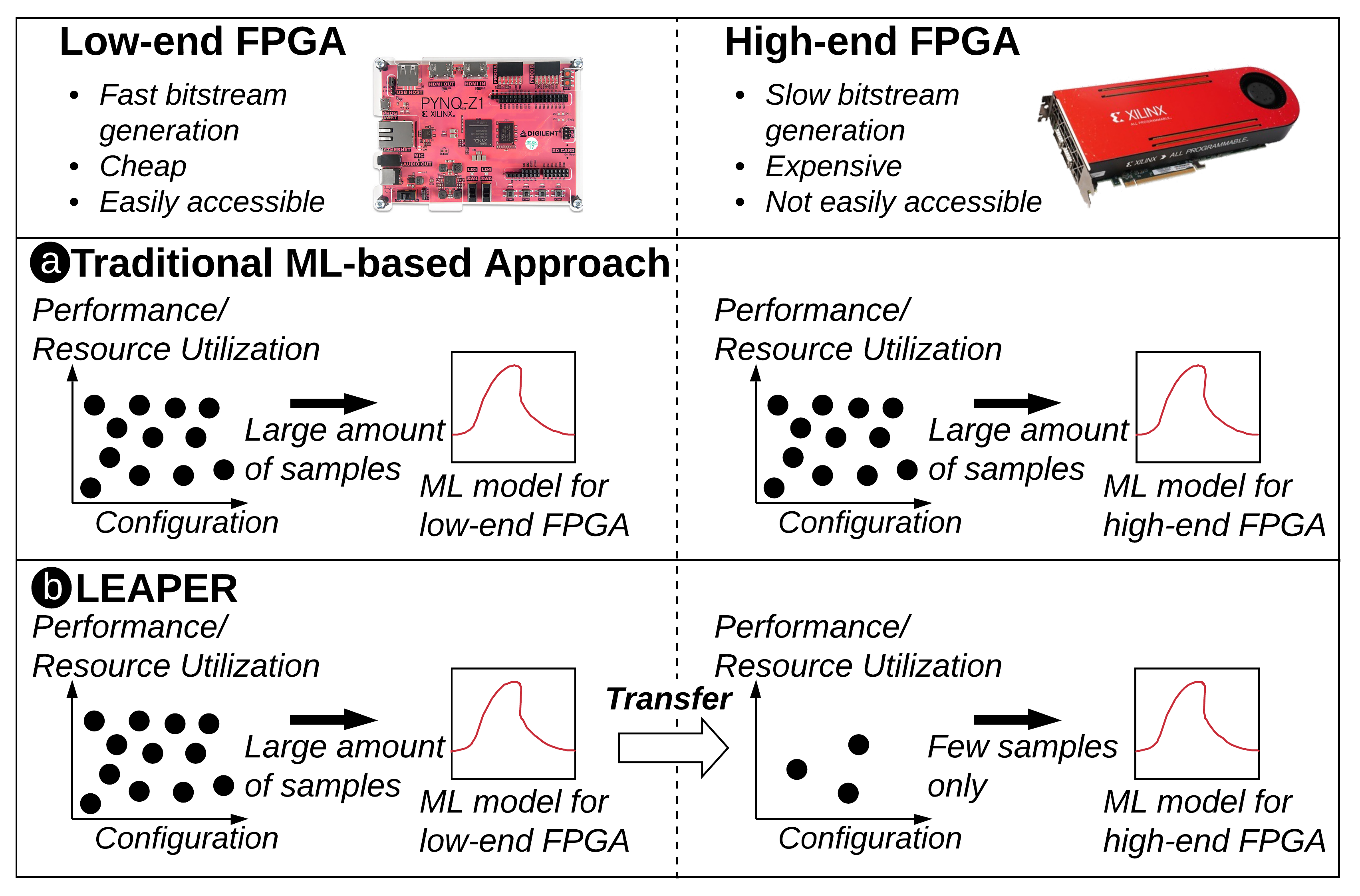}
  \caption{Traditional ML-based approach vs. \namePaper.} 
  \label{fig:transfer_intro}
 \end{figure}

Figure~\ref{fig:transfer_intro} compares the traditional ML-based approach (\circled{a}) to \namePaper  (\circled{b}). Using the traditional ML-based approach, we would 
\juan{need} to create two separate prediction models, \juan{one} for the \juan{low-end} edge \juan{environment} and \juan{another one for the high-end} cloud environment, each \juan{one} requiring a large number of samples. \namePaper transfers a previously trained model to a new, unknown environment using transfer learning with only a few training samples. 

 \vspace{0.1cm}
\gs{
\head{Key results} We demonstrate \omm{\namePaper} across \textit{five} {state-of-the-art,} high-end, cloud FPGA-based platforms with \textit{three} different {interconnect} technologies on \textit{six} real-world applications. 
We \omm{present} two keys results.
First, \namePaper achieves, on average \omm{across six evaluated workloads and five FPGAs, 85\% accuracy when we use use our transferred model for prediction in a cloud environment}. Second, \namePaper~\ommm{greatly} reduces (up to 10$\times$) the training overhead by \omm{transferring} a \textit{base model}, trained 
for a low-end, {edge} {FPGA} platform, \om{to predict performance or resource \ommm{usage} for an accelerator implementation} on a {new}, unknown high-end environment rather than building a new model from scratch. 
}

This work makes the following \textbf{major contributions}:

\begin{enumerate}

\item We \ommm{introduce} \namePaper, \omm{a \ommm{new} transfer learning-based framework for prediction of performance and resource \ommm{usage} in FPGA-based systems.} 

\item Unlike state-of-the-art works~\cite{makrani2019pyramid,ferianc2020improving,ustun2020accurate,o2018hlspredict,xppe_asp_dac,dai2018fast,zhao2019machine}  in FPGA modeling that use deep neural networks, we show that classic non-neural network-based models are enough to build an accurate predictor to evaluate an accelerator implementation on an FPGA.


\item We conduct an in-depth evaluation of \namePaper on real cloud systems with
various FPGA configurations, showing that \namePaper can develop cheaper, faster, and highly accurate models.
\end{enumerate}

\section{\namePaper}
\label{sec:methodology}

\namePaper~is a performance 
{and} resource estimation 
{framework} to \textit{transfer} ML models~\cite{swain1977decision,breiman2001random} across: \gs{(1) different FPGA-based platforms for a single application, and (2) different applications on the same platform.} In this section, we describe the main components of the framework.
First, we give an overview of \namePaper~(Section~\ref{subsec:overview}). Second, we describe the target cloud FPGA-based platform (Section~\ref{subsec:system}). Third, we discuss  accelerator optimization options and application features used for training an ML model (Section~\ref{subsec:fpga_deploy}). Fourth, we briefly describe the {base} model building (Section~\ref{subsec:base_model}).  Fifth, we explain the 
 {key} component of \namePaper to build cloud models:
 the transfer learning technique (Section~\ref{subsec:ensemble_transfer}). Sixth, we describe \namePaper's transfer learning algorithm (Section~\ref{subsec:transfer_algo}).
 
\subsection{Overview}
\label{subsec:overview}
\begin{figure}[h]
  \centering
  \includegraphics[width=1\linewidth,trim={0.2cm 0.8cm 0.4cm 0.2cm},clip]{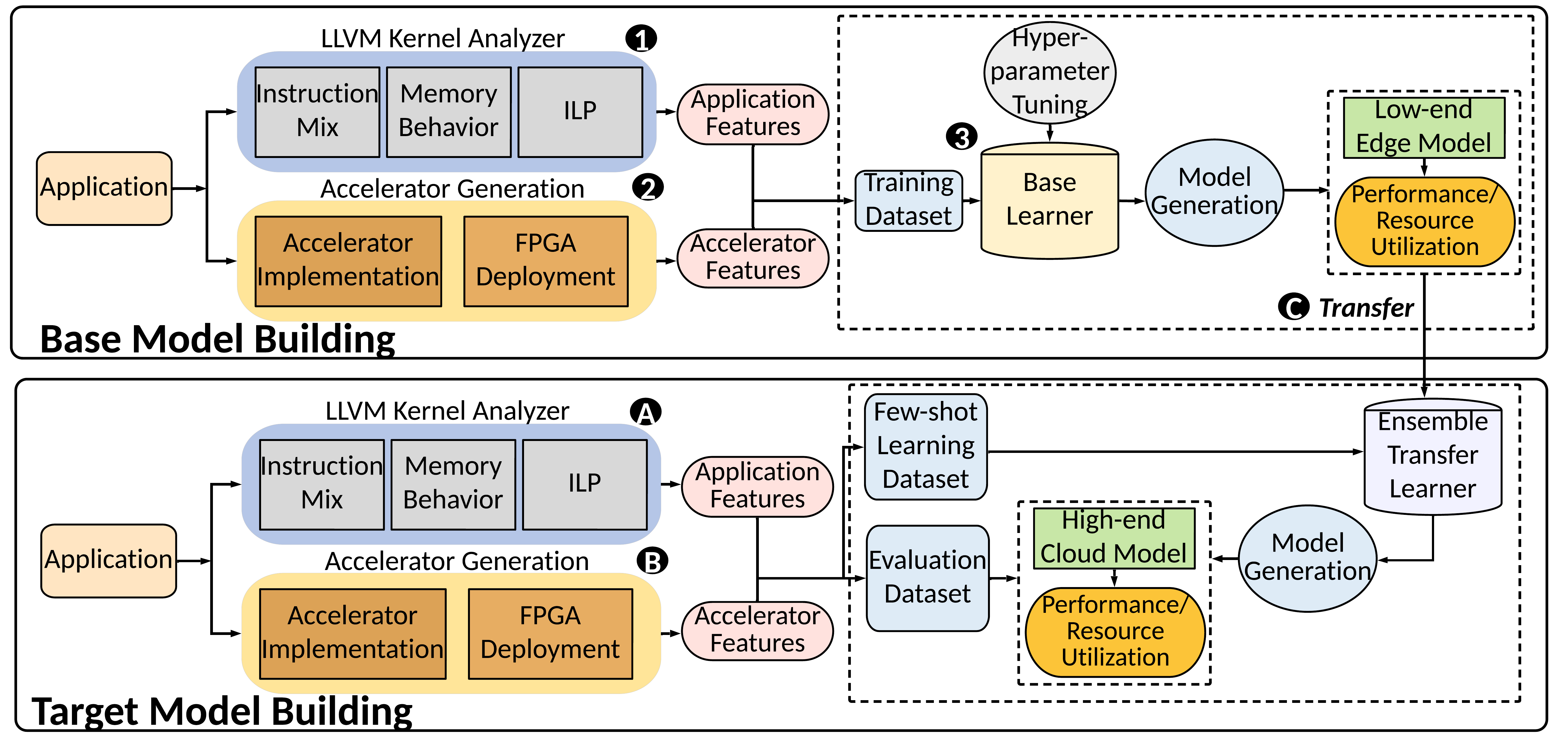}
  \caption{Overview of 
 \namePaper. 
}
  \label{fig:transfer_overiew}
 \end{figure}

Figure~\ref{fig:transfer_overiew} \gs{shows} the key components of \namePaper 
. It consists of two parts: (1) base model building (low-end environment) and (2) target model building (high-end environment). 

\noindent \textbf{Base model building.} \namePaper's base model building consists of three phases. The first phase (\circled{1} in Figure~\ref{fig:transfer_overiew}) is an LLVM-based~\cite{llvm} kernel {analysis} phase (Section~\ref{subsec:fpga_deploy}), which extracts architecture-independent workload characteristics.  We characterize in a microarchitecture-independent manner by using a specialized plugin of the LLVM compiler framework~\cite{Anghel2016}. This type of  characterization excludes any hardware dependence and captures the inherent {characteristics} of workloads.

In the second phase~\circled{2}, we generate accelerator implementations to gather accelerator \omu{implementation responses \ommm{(performance and resource utilization)}} for training. Once the accelerator design has been implemented, the {resulting FPGA-based} accelerator is deployed {in a system} with a host CPU. 
We employ the \emph{design of experiments}  (DoE) technique~\cite{montgomery2017design}
to select a small set of accelerator optimization configurations that well represent the entire space of accelerator optimization configurations ($c_{doe}$) to build a highly accurate \textit{base learner}. 
\juan{By using DoE, we minimize} the number of experiments needed to gather training data for \namePaper~while ensuring {good} quality training data. 
{Then,} we 
{run} {the} $c_{doe}$ configurations on 
{the deployed FPGA-based system} to gather samples 
for training our base model. The generated responses along with application properties from the first phase and the accelerator optimization parameters form the input to our base learner.
In the third phase~\circled{3}, we train our base learner (Section~\ref{subsec:base_model}) using ensemble learning~\cite{opitz1999popular}. 
During this phase, we perform additional tuning of our ML algorithm's hyper-parameters.\footnote{Hyper-parameters are sets of ML algorithm variables that can be tuned to optimize the accuracy of the prediction model.}  
Once trained, the framework can predict the performance and resource usage {on the base environment (a low-end edge system)} of 
{previously-unseen} configurations, 
{which} are not part of the $c_{doe}$ {configurations used during the training.} 

\begin{figure*}[h]
\centering
\begin{subfigure}[h]{.45\textwidth}
  \centering
  \includegraphics[width=\linewidth]{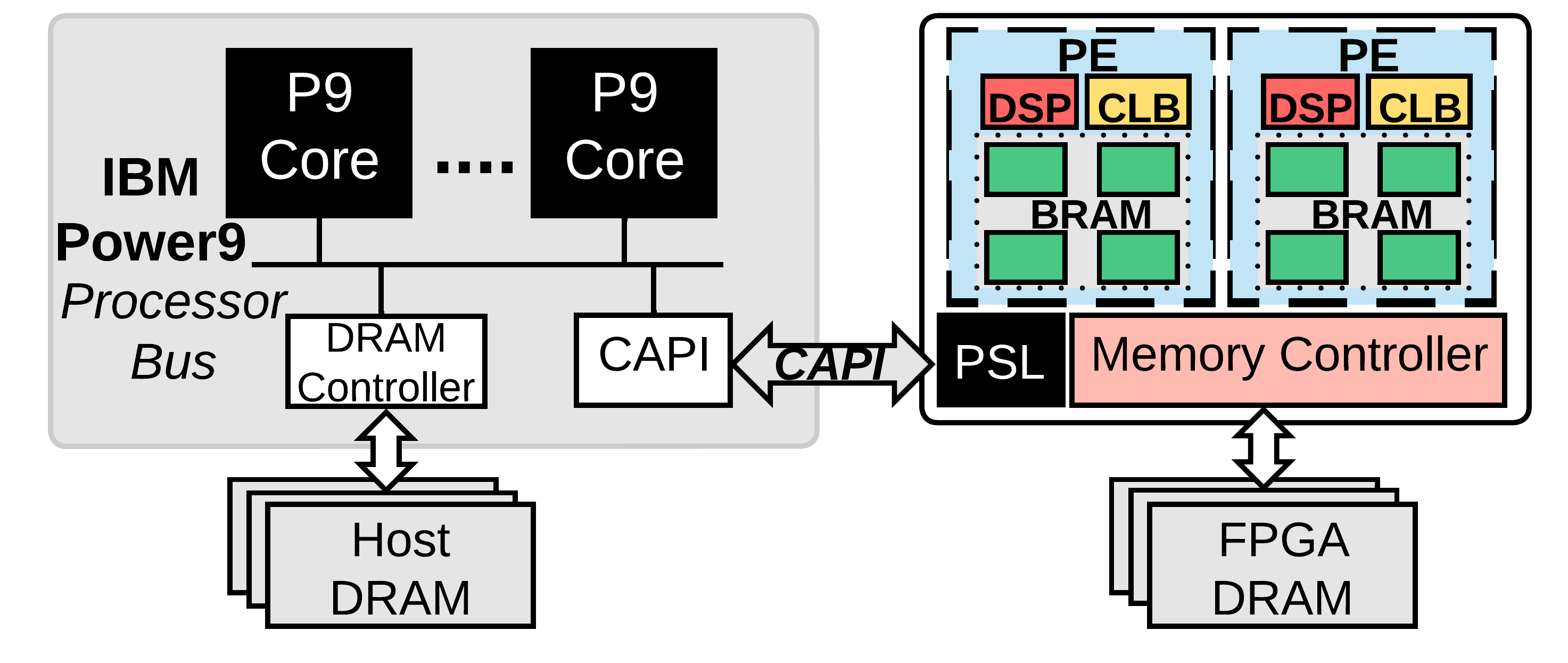}
  \vspace{-1.2cm}
  \caption{
  \label{fig:system}}
\end{subfigure}%
\hspace{0.7cm}
\begin{subfigure}[h]{.3\textwidth}
  \centering
  \includegraphics[width=\linewidth,trim={0cm 0cm 0cm 0cm},clip]{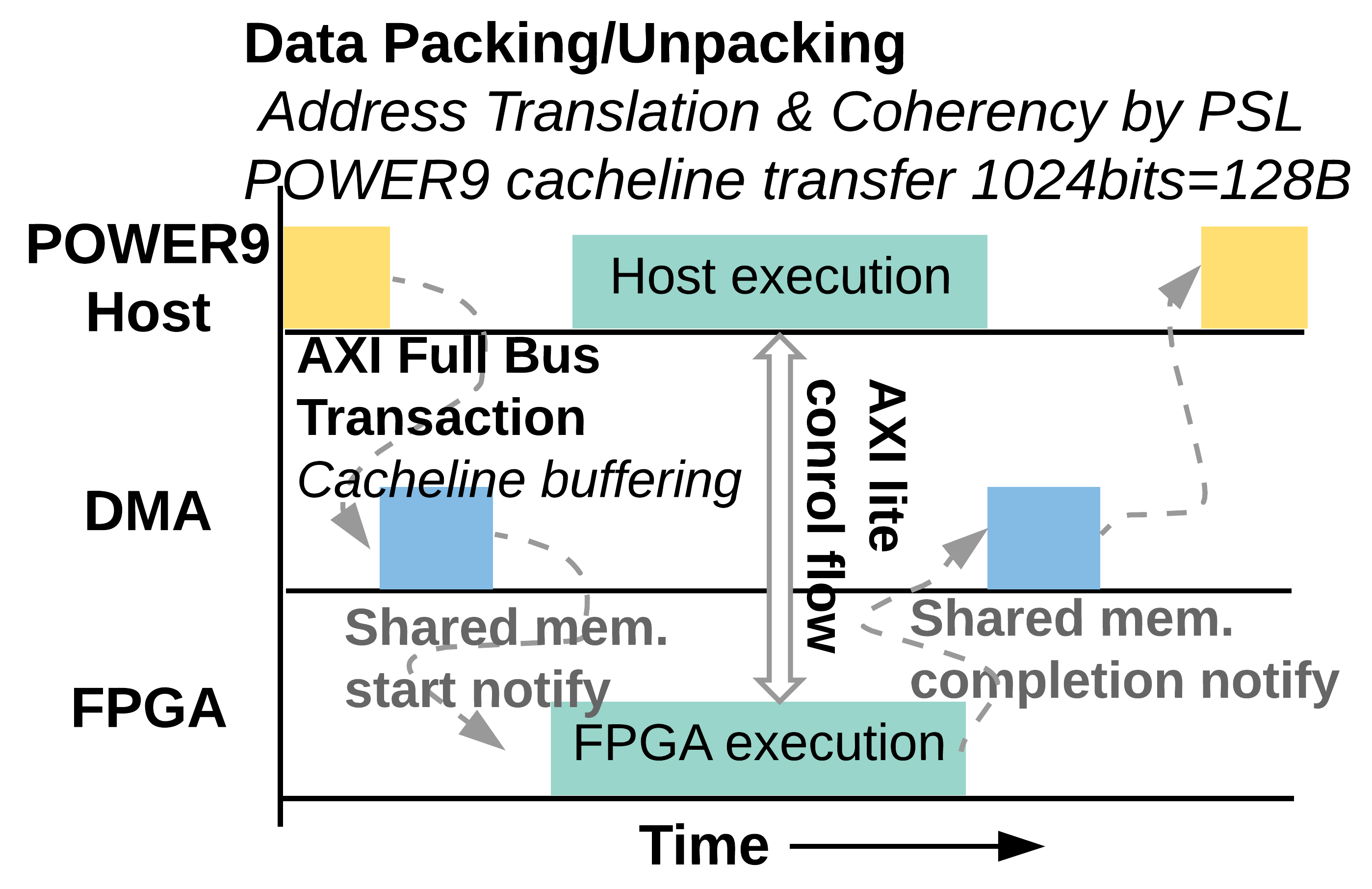}
\vspace{-1.2cm}
   \caption{ \hspace{-6.2cm}
  \label{fig:execution}}
\end{subfigure}
\caption{(a) Experimental cloud platform: High-end cloud FPGA platform with IBM\textsuperscript{\textregistered} POWER9  CPU. 
(b)  Execution timeline with data flow sequence from the  host to the FPGA.} 
\end{figure*}
\noindent \textbf{Target model building.} 
\gs{\namePaper's target model building consists of three phases.} 
The first and the second phase\ommm{s} (\circled{A} and \circled{B} {in Figure~\ref{fig:transfer_overiew}}) \ommm{are} the same LLVM-based kernel analysis and the accelerator generation phases as in base model building, respectively. 
We perform this step to create our \textit{few-shot}  learning dataset ($c_{tl}$), 
 which is used to adapt the low-end edge model to the target cloud environment.\footnote{We show \ommm{via} experiments (Section~\ref{sec:result}) that up to 5 samples (\emph{5-shot}) are enough to learn the \ommm{characteristics} of a new environment.} 
In the final phase \circled{C}, we train our ensemble transfer learner (Section~\ref{subsec:ensemble_transfer}) to leverage the low-end edge model to perform predictions for a new target environment (new application or FPGA). 

\subsection{Target Cloud FPGA-based Platform}
\label{subsec:system}

\gs{Figure~\ref{fig:system} shows the high-level overview of our  FPGA-based cloud  platform. An FPGA is connected to a server system
based on an IBM POWER9 processor using the Coherent Accelerator Processor Interface  (CAPI)~\cite{stuecheli2015capi}. Our
design consists of multiple processing elements (PEs) that
interact with the host system through the power service
layer (PSL), which is the CAPI endpoint on the FPGA. A PE can utilize the on-chip FPGA elements such as  DSP (digital signal processor), CLB (configurable logic block)\footnote{CLB is the fundamental component of an FPGA, made up of look-up-tables (LUTs) and flip-flops (FF).}, and BRAM (block RAM) to implement an application.}

Figure~\ref{fig:execution} shows the execution timeline from our host CPU to the FPGA board. We make use of CAPI in a coarse-grained way as we offload the entire application to the FPGA. CAPI ensures that a PE accesses the entire CPU memory with the minimum number of memory copies between the host and the FPGA, e.g., avoiding the intermediate buffer copies that  traditional PCIe-based DMA invokes~\cite{10.1145/2897937.2897972,singh2020nero}. However, depending on the application, the CAPI protocol can be employed in a fine-grained algorithm-hardware co-design, like in  \textit{ExtraV} \cite{extraVLee2018VLDB}, \omu{which exploits} the fine-grained communication capability of CAPI. On task completion, the PE notifies the host CPU via the AXI lite interface~\cite{axilite} and transfers back the results via CAPI-supported DMA transactions.

\subsection{Application Features and Accelerator  Optimization Options}
\label{subsec:fpga_deploy}
The ML feature vector used for training our base and target models is composed of two parts: application features and accelerator optimization options.

\noindent \juan{\textbf{Application features.}}
{We include inherent application features in our training dataset.}
{For each application kernel \textit{k} processing a dataset \textit{d}, we obtain an application profile \textit{p(k, d)}.} 
\textit{p(k, d)} is a vector where each parameter is a statistic about an application feature. 
Table~\ref{tab:application_features} lists the main application features {we extract} {by using the LLVM-based PISA analysis tool~\cite{Anghel2016}}. Ultimately, the application profile \textit{p} has 395 features, which includes all the sub-features 
{of each metric we consider}.
{We perform feature selection to select} the 100 most important features to analyze the behavior of an application. 



\begin{table}[h]
\centering
\caption{Main application features extracted from LLVM.}
\label{tab:application_features}
\centering
\small
 \renewcommand{\arraystretch}{1}
\renewcommand{\baselinestretch}{0.01}
\setlength{\tabcolsep}{10pt}
  \resizebox{1\linewidth}{!}{%
\begin{tabular}{p{0.32\linewidth} p{0.68\linewidth}}
\hline
\textbf{Application Feature} & \textbf{Description} \\ 
\hline
Instruction Mix & Fraction of \ommm{each} instruction type (integer, floating point, memory read, memory write, etc.) \\
ILP                      & Instruction-level parallelism on an ideal machine. \\
Data/Instruction Reuse Distance   & For a given distance $\delta$, probability of reusing one data element/instruction (in a certain memory location) before accessing $\delta$ other unique data elements/instructions (in different memory locations). \\
Register Traffic    & Average number of registers {per} instruction. \\
Memory Footprint    & Total memory size used by the application. \\ 
\hline

\end{tabular}
}
\end{table}


\noindent \juan{\textbf{Accelerator optimization options.}}
Table~\ref{tab:pragma} describes commonly used high-level synthesis (HLS)~\cite{hls} pragmas to optimize an accelerator design on an FPGA. These optimization options constitute a part of our ML feature vector for training. 
We use eight optimization options. First, \textit{loop pipelining} {(PL)} optimizes a loop to overlap different loop operations. Second,  \textit{loop unrolling} {(UR)} creates multiple copies of a loop for parallel execution. PL and UR aim to increase the processing throughput of an accelerator implementation.
Third, to enable simultaneous memory accesses, \textit{array partitioning} {(PR)} divides \ommm{an} array into smaller units of \ommm{desired partitioning factor} 
and maps {them} 
to 
{different memory banks.} This optimization produces considerable speedups but consumes more resources. Fourth, \textit{inlining} {(IL)} ensures that a function is instantiated as dedicated hardware core. Fifth,  \textit{dataflow} {(DF)} optimization exploits task-level parallelism to allow parallel execution of tasks. 
Sixth, \textit{burst read} {(BR)} controls burst reads from the host to the accelerator. Seventh, \textit{burst writes} {(BW)} controls burst write to the host from an accelerator. Eighth, \textit{FPGA frequency} {(FR)} constrains an accelerator implementation to a specific clock frequency. It affects {not only} the performance 
{but also} the resource \ommm{usage of} an implementation. For instance, to meet the FPGA timing requirements, the FPGA tool inserts registers between flip-flops, {which increases  resource \ommm{usage}}. 

\begin{table}[h]
\centering
  \caption{\gaganT{Accelerator optimization options used in training.}}
    \label{tab:pragma}
        \resizebox{0.85\linewidth}{!}{%
\begin{tabular}{@{}lll@{}}
\hline
Optimization &  Description  \\ \hline
Loop pipelining (PL)       &   Enabled/disabled \\
Loop unrolling  (UR)     &   Unrolling factor (Factor: $2^n, 1\leq n \leq 6$) \\
Array partitioning (PR)   &  Block/cyclic/complete (Factor: $2^n, 1\leq n \leq 6$)) \\

Inlining (IL)        &   Enabled/disabled function inlining \\
Dataflow (DF)        &  Task level pipelining \\
Burst read (BR)     &   Read data burst from the host \\
Burst write (BW)    &   Write data burst \ommm{to} the host \\

FPGA frequency (FR)     &  Four-different frequency levels for an FPGA logic\\ & (50 MHz, 100 MHz, 150 MHz, and 200 MHz)   \\
\hline
\end{tabular}
}
\end{table}

In total, our optimization options {for a particular application} leads {up to} 4,608 configurations. However, the actual optimization space of an application depends 
on the specific application characteristics (see Table~\ref{tab:app_details}). 
{For example, we include} loop unrolling {in the optimization space when an application contains loops that can be unrolled.}  

\subsection{Base Learner Training}
\label{subsec:base_model}

The third phase of base model training is the training of the base learners. We use {an \emph{ensemble} of} two \gs{non-linear} base learners \gs{that {can} capture the intricacies of accelerator implementation by predicting the execution time 
or the resource \ommm{usage}.} 
Our first algorithm is the \emph{random forest} (RF)~\cite{breiman2001random}, which is based on \emph{bagging}~\cite{breiman1996bagging}. 
We use RF to avoid a complex feature-selection scheme 
{since RF} embeds automatic procedures {that are able} to screen many input features~\cite{mariani2017predicting}. 
Starting from a root node, RF constructs a tree and iteratively grows the tree by associating 
{a node} with a splitting value for an input 
{feature} to generate two child nodes. 
Each node is associated with a prediction of the target metric equal to the observed {mean} value in the training dataset for the input subspace that the node represents. 
Our second learner is \emph{gradient boosting}~\cite{gradientboosting}. 
\gs{Gradient boosting aims to \textit{boost} {the accuracy of a weak learner} by using other learners to correct its predictions.}
Bagging~\cite{schapire1990strength} reduces model variance, and boosting decreases errors~\cite{kotsiantis2004combining}. Therefore, we use RF and gradient boosting together to increase the predictive power of our final trained base model.
In a machine learning task, $\mathcal{X}$ represents \ommm{the} feature space with label $\mathcal{Y}$, where a machine learning model is responsible for estimating a function $f: \mathcal{X} \to \mathcal{Y}$. 
{\namePaper uses \textit{base learners} to predict the performance (or resource \ommm{usage}) $\mathcal{Y}$ for a tuple} \textit{(p, c)} {that belongs to the ML feature space $\mathcal{X}$}, where \textit{c} is \ommm{a set of} accelerator optimization options  
\ommm{for an} application profile \textit{p(k,~d)}.

{The training dataset for our base model} has three parts: (1)  an application configuration vector \textit{p(k,~d)}, (2) an accelerator optimization option 
\textit{c}, and (3) responses corresponding to each pair \textit{(p, c)}.
To gather the accelerator responses, {we run each} application \textit{k} belonging to {the} training set $\mathbb{T}$ with {an} input dataset \textit{d} 
on an FPGA-based platform \ommm{while using an accelerator optimization \textit{c}}. 
{This way, we obtain the execution time for the tuple \textit{(p, c)}, which we can use} 
as a \textit{label} ($\mathcal{Y}$) for training our base learner {for performance prediction}. 
{We build a similar model} to predict resource \ommm{usage}, where we use {the} resource \ommm{usage} ($\eta_{\{BRAM, FF, LUT,  DSP\}}$) {of the tuple \textit{(p(k,~d), c)}} as a \textit{label} when we train our base learner {for resource \ommm{usage}}. 
After \ommm{it is trained,} our base learner 
it can predict the {execution} time (or resource usage) ($\hat{f_s}:\mathcal{X}_s \to \mathcal{Y}_s$) of 
{tuples \textit{(p(k,~d), c)}} that are \emph{not} in the training set.

We improve base learner performance by tuning the algorithm’s hyper-parameters~\cite{scikit-learn}. Hyper-parameter tuning can provide better performance estimates for some applications. First, we perform as
many iterations of the cross-validation process as hyper-parameter
combinations. Second, we compare all the generated models by
evaluating them on the testing set, and select the best one.

\subsection{{{Target} Model Building via Transfer Learning}}
\label{subsec:ensemble_transfer}
{
\namePaper provides 
the ability to transfer trained \ommm{a} performance (or resource \ommm{usage}) model across different environments.}
{\namePaper~defines} a target environment $\tau_{t}$ as an environment for which we wish to build a prediction model {$\hat{f}_t$ where data collection is expensive}, and 
a source environment $\tau_{s}$ as an environment 
 for which we can \textit{cheaply} collect many samples to build an ML model ${f}_s$. {In our case, $\tau_{s}$ is a low-cost edge FPGA-based system, while $\tau_{t}$ is a high-cost cloud FPGA-based system. } {\namePaper~then transfers the ML model for $\tau_{s}$ to $\tau_{t}$ using \ommm{an} ensemble transfer model $\hat{h}_t$.}

\head{Transfer learner} In transfer learning, a weak relationship between the base and the target environment can decrease the predictive power of the target environment model. This degradation is referred to as a \textit{negative transfer}~\cite{jamshidi2017transfer}. To avoid this, we use an ensemble 
model trained on the transfer set (i.e., {the} \textit{few-shot learning} dataset in Figure~\ref{fig:transfer_overiew}) as our transfer learners (TLs). 
We use non-linear transfer learners because, based on our analysis ({Section}~\ref{section:relatedness_analysis}), non-linear models are able to \gs{better} capture the non-linearity present in the accelerator performance and optimization options. Our first TL is based on TrAdaBoost~\cite{dai2007boosting},  a boosting algorithm that fuses many weak learners into one strong predictor by adjusting the weights of training instances. {The motivation behind such an approach is that by fusing many weak learners, boosting can improve the overall predictions
in areas where the \ommm{individual} learners did not perform well.} We use Gaussian process regression~\cite{Gaussian} as our second TL. It is a Bayesian algorithm that calculates the probability distribution over all the appropriate functions that fit the data.  To transfer a trained model, we train both TrAdaBoost and Gaussian process regresssion, and select the best performing TL.

\subsection{\namePaper Transfer Learning Algorithm}
\label{subsec:transfer_algo}
Algorithm~\ref{algo:transfer} presents \namePaper's transfer learning approach. \gs{We provide \ommm{as} input the: (1) ${f}_s$ model trained for a low-end environment, and (2) sub-sampled few-shot learning dataset \textit{c$_{tl}$}.}  
\gs{We initialize the training loop to the maximum value (line~\ref{algo:tl_initial}) to run \namePaper until convergence  (line~\ref{algo:tl_initial_loop}). We normalize the input feature vector to have all the different features to be on the same scale (line~\ref{algo:tl_data}).
Using the normalized input data, \namePaper~ trains the ensemble of TLs (line~\ref{algo:tl_train}) that} transforms the performance or resource \ommm{usage} model of a source environment  $f_{s}$ to the target environment's performance or resource \ommm{usage} model $f_{t}$. 
\ommm{We use \textit{c$_{tl}$} to generate a transfer learner $\hat{h}_t$ (line~\ref{algo:line:findTL}). We choose the TL that has the lowest mean relative error (line~\ref{algo:tl_mre}). Finally, we use $\hat{h}_t$ to transfer predictions from ${f}_s$ to produce ${f}_t$ (line~\ref{algo:tl_transform}) by performing a non-linear transformation (line~\ref{algo:tl_transfer_predict}).} 

\begin{algorithm}[h]
 \caption{\namePaper's~transfer learning \gs{algorithm.}}
\setstretch{0.55}
\small
 \label{algo:transfer}
\SetAlgoLined\DontPrintSemicolon
 \SetKwInOut{Input}{Input}
\SetKwInOut{Output}{Output}
  \KwIn{(1) Base model ($f_s$) trained \ommm{on} the edge environment,\\  \hskip3em (2) Sub-sampled {\emph{few-shot learning} dataset $c_{tl}\subset c_{doe}$\\ \hskip3em from the base and the target environment}}   
  \KwOut{Target {cloud} model ${f}_t:\mathcal{X}_t \to \mathcal{Y}_t$}
\textbf{Initialize:} Maximum number of iterations M\label{algo:tl_initial}\;
\While{$M\not=0$}{\label{algo:tl_initial_loop}
    Normalize the feature vector\label{algo:tl_data}\;
    Train ensemble transfer learners (TL) with $c_{tl}$\label{algo:tl_train}\;
   Find the candidate TL:\;\hskip1em  $\hat{h}_t:\mathcal{X}_{tl} \to \mathcal{Y}_{tl}$ that minimizes the error over the $c_{doe}-c_{tl}$\; \label{algo:line:findTL}
    Compute the mean relative error\label{algo:tl_mre}: \\
   \hskip1em $\epsilon_{mre}=\frac{1}{c_{doe}-c_{tl}} \displaystyle\sum_{i=1}^{c_{doe}-c_{tl}} \frac{|y_{t}^{acc} -y_{t}^{pred}|}{y_{t}^{acc}}$
   
    Use identified $\hat{h}_t$ to transform predictions of $f_s$:\label{algo:tl_transform}\\
   \hskip1em${f}_t$=$\hat{h}_t(f_s)$  \label{algo:tl_transfer_predict} \hskip7em where $f_s: \mathcal{X}_s \to \mathcal{Y}_s$\\
   $M \leftarrow M - 1$
}
\textbf{return}  ${f}_t$
\end{algorithm}



\section{{Evaluation \gs{Methodology}}}
\label{sec:evaluation}

\subsection{Hardware Platform}
\label{subsec:eval_platform}
Table~\ref{tab:systemparameters} summarizes the system details of our low-end edge environment and {high-end cloud} environment. We select \ommm{the} widely available PYNQ-Z1 board~\cite{pynq} with XC7Z020-1CLG400C FPGA~\cite{zynq} as the low-end FPGA platform to build base model. We use the accelerator coherency port (ACP)~\cite{ACP} with PCIe Gen2  \ommm{to} attach FPGA-based accelerators to the ARM Cortex-A9 CPU~\cite{cortexA9} present \ommm{in} PYNQ-Z1. 
We use \ommm{the} Nimbix cloud~\cite{nimbix} with CAPI-based FPGA system attached to a server-grade IBM POWER9 CPU system as our high-end cloud environment. \ommm{Nimbix} uses KVM (Kernel Virtual Machine)~\cite{kvm} for Linux virtualization and OpenStack~\cite{openstack} as  middleware. We evaluate five state-of-the-art, high-end, cloud FPGA boards (ADM8K5~\cite{ad8k5}, ADM9V3~\cite{ad9v3}, NSA241~\cite{nsa241}, N250SP~\cite{250SP}, and ADMKU3~\cite{adku3}) using two different interconnect technologies (CAPI-1 and CAPI-2). We can derive from the indicative prices listed in the last column of  Table~\ref{tab:systemparameters} \ommm{shows} that the  total cost of ownership (TCO) of a high-end cloud system can be more than $100\times$ of that of the low-end system, and thus it can be cost-prohibitive 
to many users.

%

\begin{table}[h]
\centering
  \caption{System parameters and configuration.}
    \label{tab:systemparameters}
    \renewcommand{\arraystretch}{0.9}
\setlength{\tabcolsep}{1pt}
    \resizebox{\linewidth}{!}{
\begin{tabular}{@{}lllllr@{}}
\toprule
\multicolumn{5}{l}{\textbf{Low-end Edge System}} & \textbf{Indicative Price}                                                                                                                                                                               \\ \midrule

    & 
\multicolumn{4}{l}{PYNQ-Z1 ZYNQ~\cite{pynq} XC7Z020-1CLG400C~\cite{zynq} with} & \$299\footnote{https://store.digilentinc.com/pynq-z1-python-productivity-for-zynq-7000-arm-fpga-soc/, Accessed 12 Jan. 2021} \\

&\multicolumn{4}{l}{PCIe Gen2~\cite{pcie} ARM Cortex-A9  @650MHz, dual-core} \\ 
&\multicolumn{4}{l}{512MB DDR3 with 16-bit bus @ 1050Mbps} \\

\toprule
\toprule
\multicolumn{5}{l}{\textbf{Nimbix Cloud~\cite{nimbix} System with OpenStack\cite{openstack} and KVM Hypervisor~\cite{kvm}}}     & \textbf{Indicative Price}                             \\ \midrule
\textbf{Host Configuration}    & 
\multicolumn{4}{l}{IBM\textsuperscript{\textregistered} POWER9 AC922~\cite{POWER9} @2.3 GHz, 16 cores} & \$55000-\$75000\footnote{https://www.microway.com/product/ibm-power-systems-ac922/} \\ 
&\multicolumn{4}{l}{4-way SMT~\cite{smt}, 32 KiB L1 cache,256 KiB L2 cache,} \\ 
&\multicolumn{4}{l}{10 MiB L3 cache, 32GiB RDIMM DDR4 2666 MHz~\cite{rdimm}} \\ 

\midrule
\textbf{FPGA Description} & \multicolumn{4}{l}{}      &                             \\ 

 \textbf{Board} &\textbf{FPGA Family}        &\textbf{Device}   &\textbf{Interface}& & \textbf{Indicative Price}      \\    
ADM9V3~\cite{ad9v3}   &Virtex UltraScale\tnote{+} & XCVU3P-2  & { CAPI-2 }   & & N/A  \\
NSA241~\cite{nsa241}& Virtex UltraScale\tnote{+} &XCVU9P-2  & { CAPI-2 }  & & N/A  \\
N250SP~\cite{250SP}&Kintex UltraScale\tnote{+} &KU15P-2  &{ CAPI-2 }  & & N/A  \\     
ADMKU3~\cite{adku3}    &Kintex UltraScale &XCKU060-2  & { CAPI-1 }  & & N/A \\
ADM8K5~\cite{ad8k5}   &Kintex UltraScale &XCKU115-2   & { CAPI-1 }  & & N/A \\
\bottomrule
\end{tabular}
}
\vspace{-0.3cm}
\begin{flushleft}
\scriptsize \textsuperscript{5} https://store.digilentinc.com/pynq-z1-python-productivity-for-zynq-7000-arm-fpga-soc/ \ommm{(accessed on 2022-06-13)}\\
\scriptsize \textsuperscript{6} https://www.microway.com/product/ibm-power-systems-ac922/ \ommm{(accessed on 2022-06-13)}\\
N/A: Not available indicative price from an online store, but in the region of \$2500-\$5000 for our purchased on-prem cards.
\end{flushleft}
\end{table}
\vspace{-0.4cm}

\subsection{Programming Toolflow}

We use the {Xilinx} SDSoC~\cite{sdsoc} design tool for implementing an accelerator on the low-end edge environment
$\tau_{s}$ and the Vivado HLS~\cite{hls} with \ommm{the} IBM CAPI-SNAP framework\footnote{https://github.com/open-power/snap} for the high-end cloud environment $\tau_{t}$. The SNAP framework provides seamless integration of an accelerator~\cite{10.1007/978-3-030-34356-9_25} and allows \ommm{the} exchange of control signals between the host and the FPGA processing elements over the
AXI lite interface~\cite{axilite}.

\begin{figure*}[h]
  \centering
  \includegraphics[width=\linewidth]{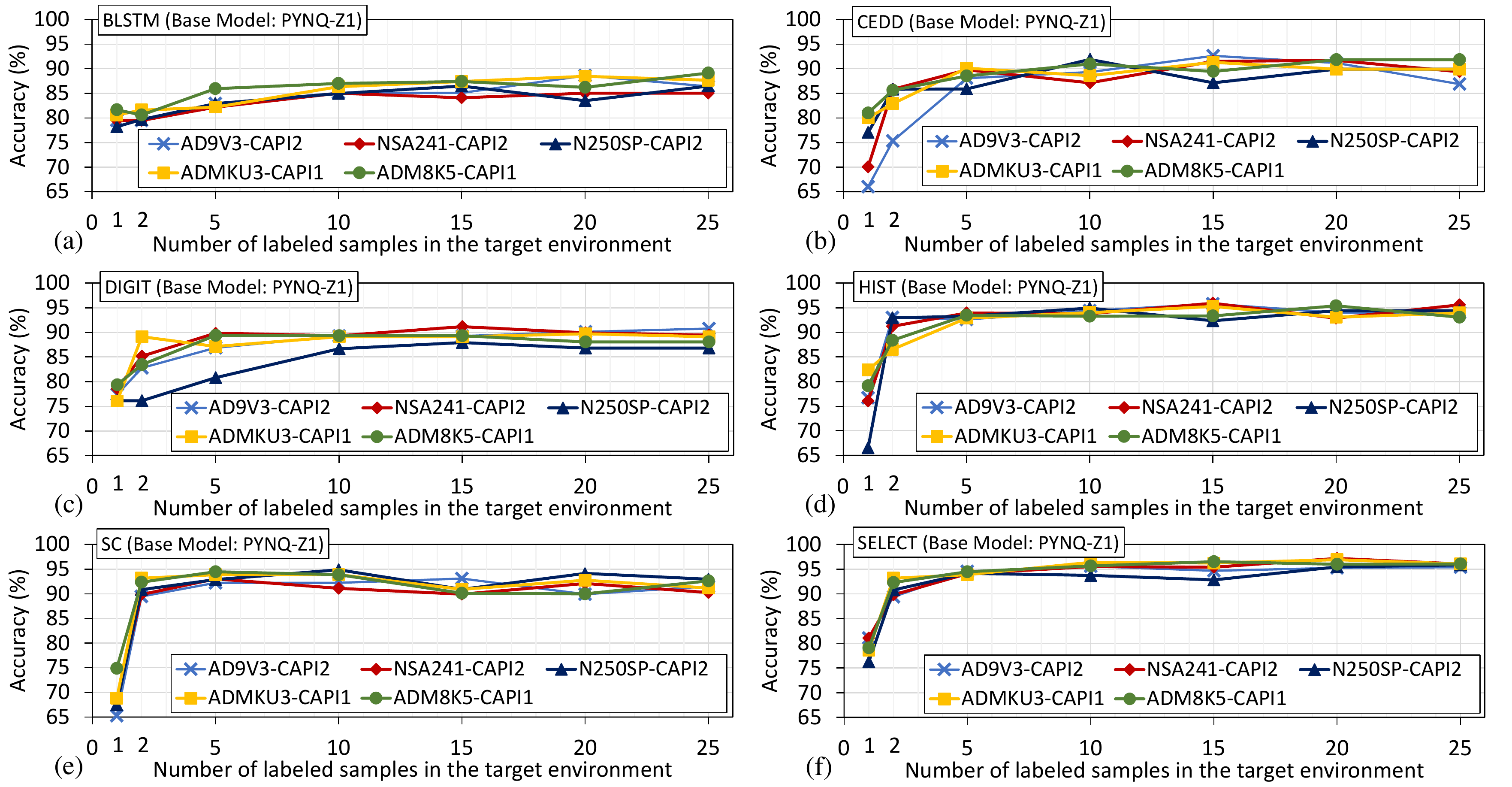}
  \vspace{-15pt}
  \caption{\namePaper's accuracy for 
  platforms using different samples {(horizontal axis)} from the target platform. }
  \label{fig:acros_board}
\end{figure*}

 \begin{figure*}[t]
  \begin{subfigure}{0.495\textwidth}
   \includegraphics[width=\textwidth,trim={0cm 0cm 0cm 0cm},clip]{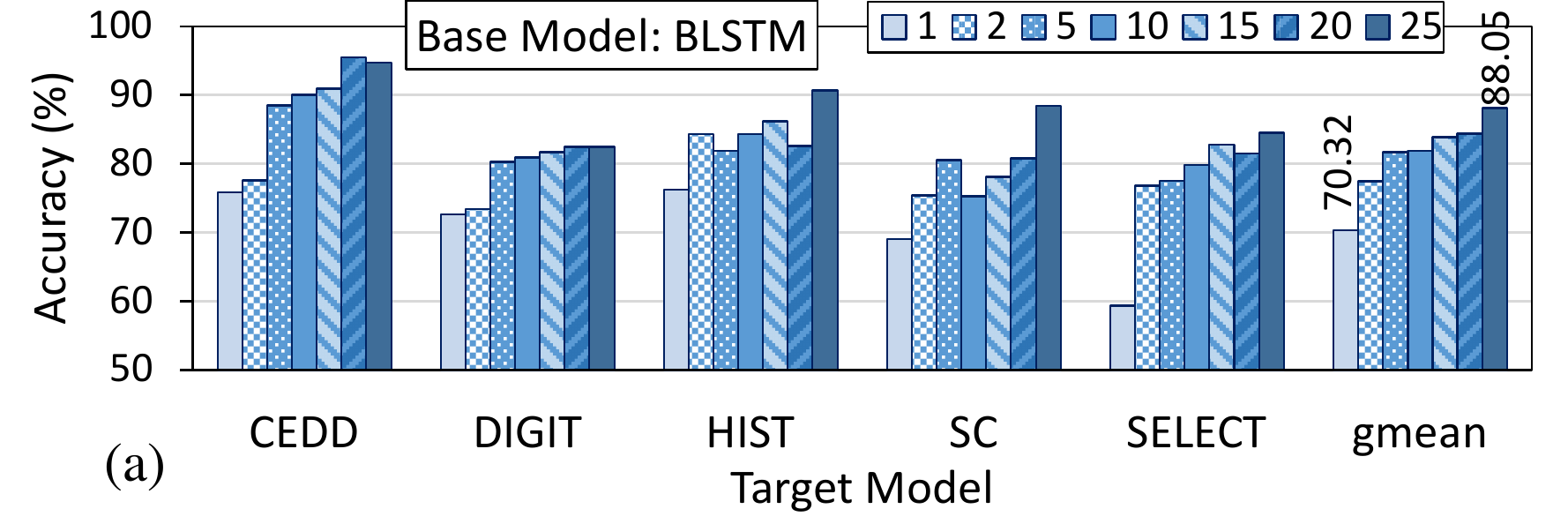}
     \vspace{-0.3cm}
    \label{fig:app_blstm}
  \end{subfigure}
  \begin{subfigure}{0.495\textwidth}
  \includegraphics[width=\textwidth,trim={0cm 0cm 0cm 0cm},clip]{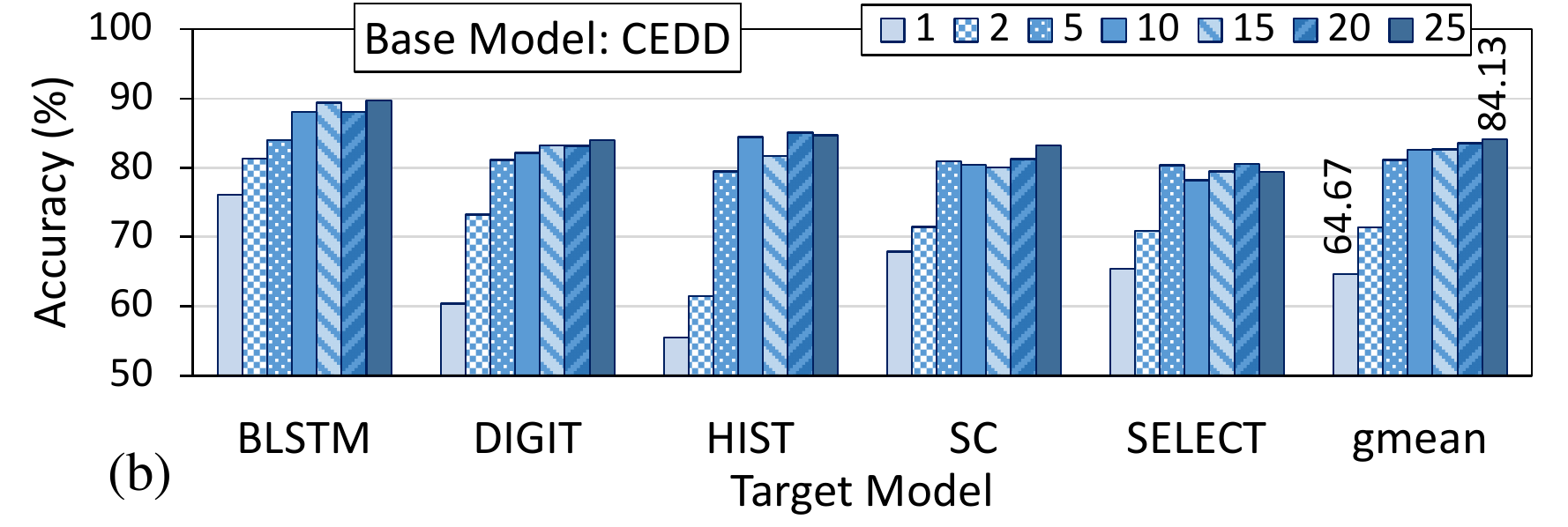}
    \vspace{-0.3cm}
    \label{fig:app_cedd}
  \end{subfigure}
   \begin{subfigure}{0.495\textwidth}
\includegraphics[width=\textwidth,trim={0cm 0cm 0cm 0cm},clip]{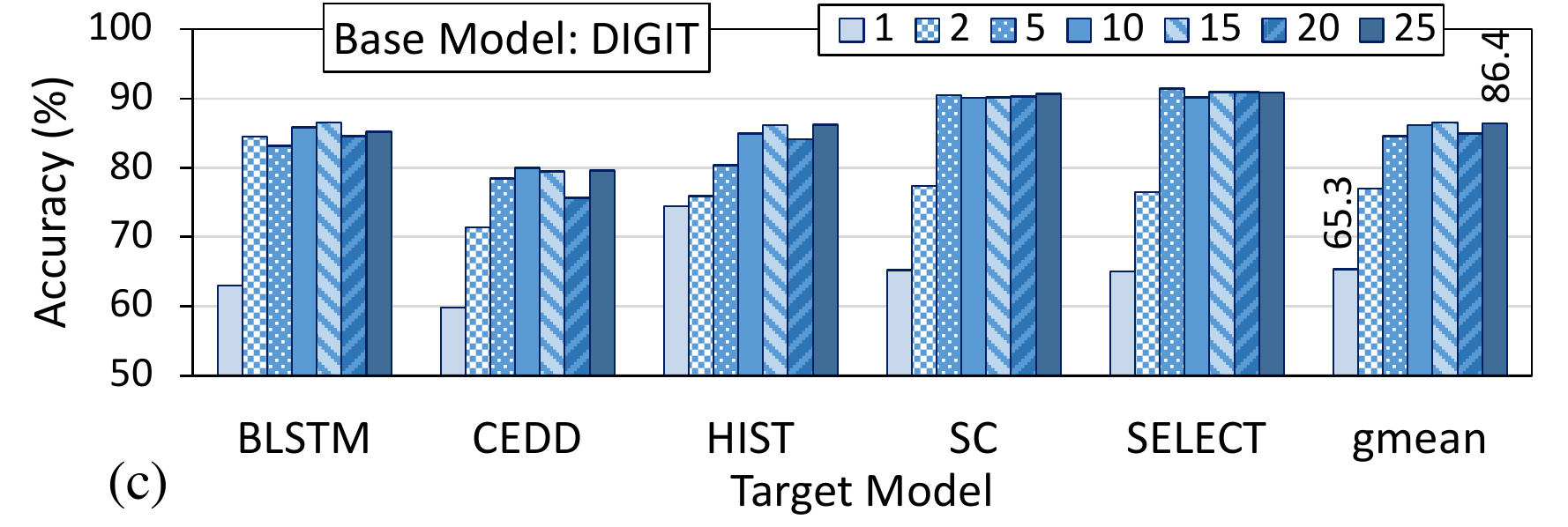}    \vspace{-0.3cm}
    \label{fig:app_digit}
  \end{subfigure}
   \begin{subfigure}{0.495\textwidth}
    \includegraphics[width=\textwidth,trim={0cm 0cm 0cm 0cm},clip]{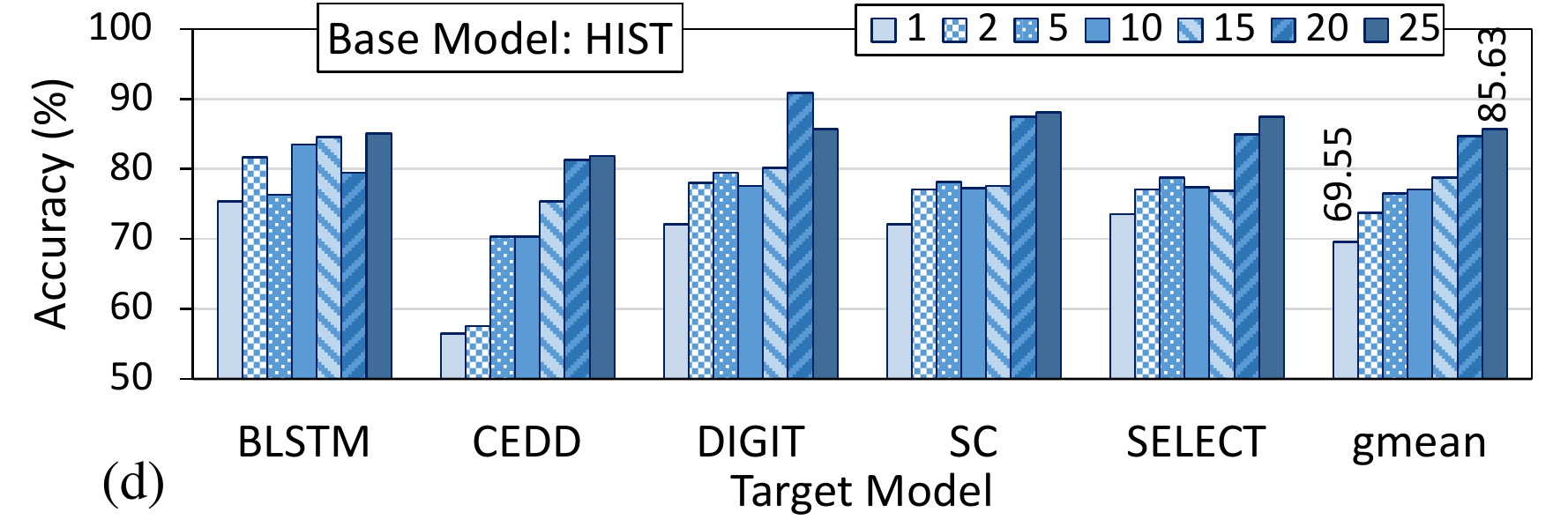}
   \vspace{-0.3cm}
    \label{fig:app_hist}
  \end{subfigure}
     \begin{subfigure}{0.495\textwidth}
    \includegraphics[width=\textwidth,trim={0cm 0cm 0cm 0cm},clip]{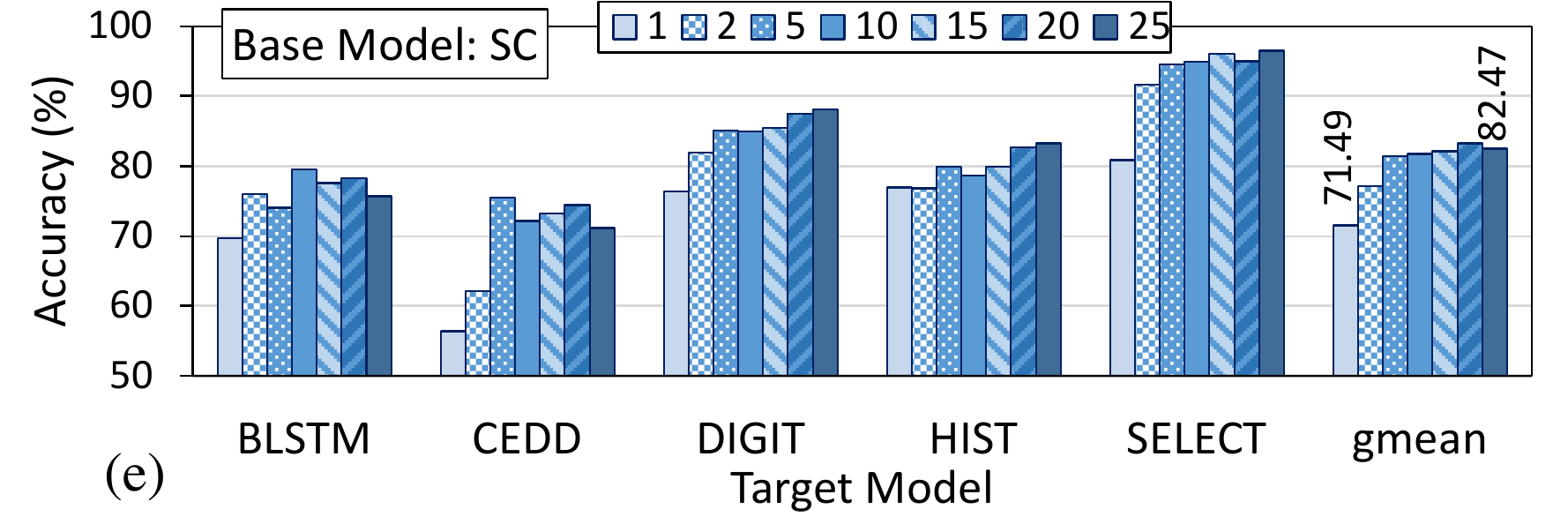}
    \label{fig:app_sc}
  \end{subfigure}
     \begin{subfigure}{0.495\textwidth}
     \includegraphics[width=\textwidth,trim={0cm 0cm 0cm 0cm},clip]{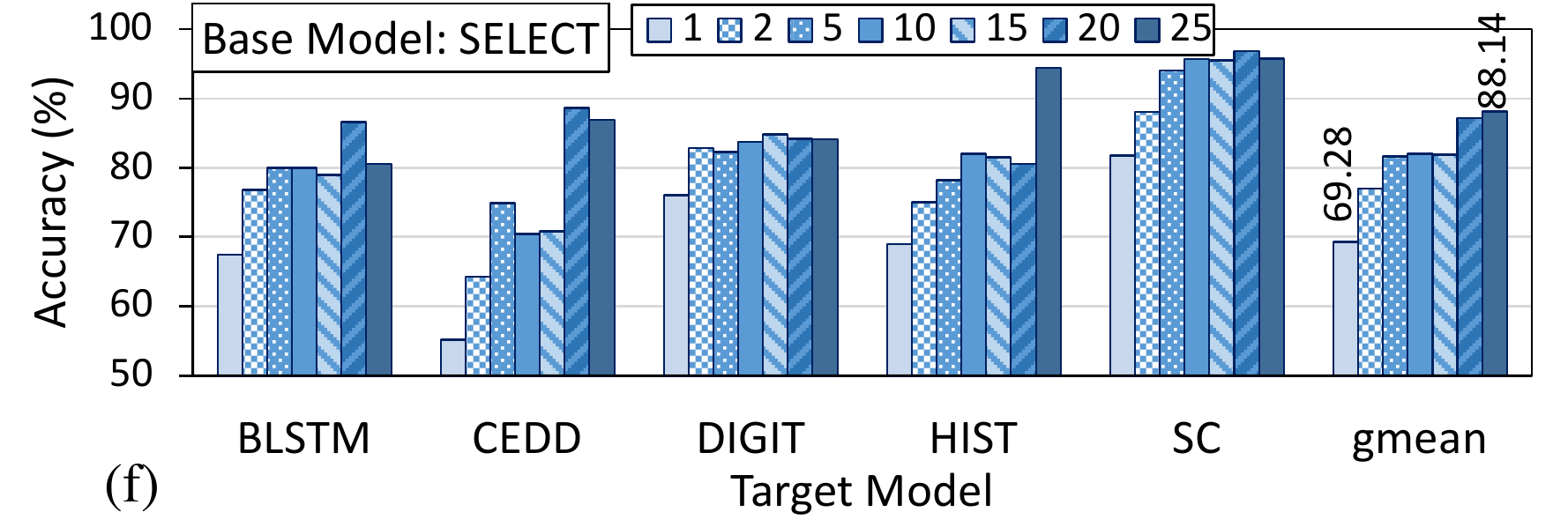}
    \label{fig:app_select}
  \end{subfigure}
\vspace{-14pt}
  \caption{\namePaper's accuracy for 
  {transferring} base models across various applications. {The legends indicate the number of samples}. Each plot represents a different application \ommm{used as a base model}. We \textit{transfer} these base models, trained on the PYNQ-Z1 platform. 
  } 
   \label{fig:across_app}
\end{figure*}


\subsection{Workloads} 
\label{subsec:eval_workload}
We evaluate \namePaper~using six
 benchmarks (Table~\ref{tab:app_details}), which are hand-tuned for FPGA execution. 
These benchmarks cover 
{several application} domains, i.e., \textbf{(1) image processing}:  histogram {calculation (\textit{hist})~\cite{gomezluna2017chai}, and canny edge detection (\textit{cedd})~\cite{gomezluna2017chai}}; \textbf{(2) machine learning}: binary long short term memory (\textit{blstm})~\cite{diamantopoulos2018ectalk}, digit recognition (\textit{digit})~\cite{zhou2018rosetta}; 
\textbf{(3) databases}: relational operation (\textit{select})~\cite{gomezluna2015ds}; 
and \textbf{(4) {data reorganization}}: stream compaction (\textit{sc})~\cite{gomezluna2017chai}. These kernels are specified in C/C++ code using high-level synthesis (HLS) that is compiled to the target FPGA device.

\vspace{0.25cm}
\begin{table}[h]
  \caption{\gagan{Evaluated application description including their domain, major kernels, and the input dataset. For major kernels, we mention the optimization space where $\times$ represents the optimization being applied to multiple loops or elements. 
  }}
    \label{tab:app_details}
  \resizebox{1\linewidth}{!}{%
\begin{tabular}{l@{\hspace{0.9\tabcolsep}}l@{\hspace{0.9\tabcolsep}}l@{\hspace{0.9\tabcolsep}}l@{\hspace{0.9\tabcolsep}}l}
\hline
  \textbf{Application}                  & \textbf{Domain} & \textbf{Major Kernels}     & \textbf{Dataset}    & \textbf{Optimization Space}                                                                     \\ \hline
    blstm~\cite{diamantopoulos2018ectalk} &  \begin{tabular}[c]{@{}l@{}} Machine \\learning\end{tabular}     & \begin{tabular}[c]{@{}l@{}}Hidden layer fw\\ Hidden layer back \\ Output layer\end{tabular} & Fraktur OCR~\cite{yousefi2015binarization} & 
    \begin{tabular}[c]{@{}l@{}}2$\times$PL, 3$\times$PR(2,4), IL, 2$\times$UR \\ 2$\times$PL, IL, 2$\times$UR  \\ PL, IL, UR, DF, BR, BW, FR\end{tabular}

\\\hline 
    
     cedd~\cite{gomezluna2017chai}             & \begin{tabular}[c]{@{}l@{}}  Image \\  proc. \end{tabular}   & \begin{tabular}[c]{@{}l@{}}
     Gaussian filter\\
     Sobel filter\\
    Suppression filter\\
    Hysteresis filter\\
    \end{tabular}            & \begin{tabular}[c]{@{}l@{}}   Frame-354$\times$626 \\ 1000 frames               \\
    \end{tabular}  &
    \begin{tabular}[c]{@{}l@{}}  
    PL, PR(2,4), IL, UR\\
    PL, PR(2,4), IL, UR\\
    PL, PR(2,4), IL, UR\\
    PL, IL, UR, DF, BR, BW, FR\\

     \end{tabular} 
    \\
\hline 
    digit~\cite{zhou2018rosetta}             & \begin{tabular}[c]{@{}l@{}}  Machine \\learning \end{tabular}    & \begin{tabular}[c]{@{}l@{}}Hamming dist.\\
    KNN voting\end{tabular}          & \begin{tabular}[c]{@{}l@{}} MNIST-18000 train \\
    2000 test  \end{tabular}        &       \begin{tabular}[c]{@{}l@{}}       
     2$\times$PL, 3$\times$PR(2,4), IL, 4$\times$UR\\
IL, BR, BW, FR\\

\end{tabular}
\\\hline

    hist~\cite{gomezluna2017chai}                      &\begin{tabular}[c]{@{}l@{}}   Image\\ proc.  \end{tabular}   & Histogram avg.     &    \begin{tabular}[c]{@{}l@{}}   Input-1536$\times$1024      \\
    Bins-256\\
        \end{tabular} & \begin{tabular}[c]{@{}l@{}} PL, PR, IL, DF,\\ BR, BW, UR, FR  \end{tabular} \\
\hline 

     sc~\cite{gomezluna2017chai}                & \begin{tabular}[c]{@{}l@{}}   Data\\ reorg.  \end{tabular}  & \begin{tabular}[c]{@{}l@{}}Count\\ Compact\end{tabular}       &                  1048576 elements    & \begin{tabular}[c]{@{}l@{}}PL, IL, DF, BR, BW, UR, FR       \end{tabular}          \\ \hline
   select~\cite{gomezluna2015ds}         & Database          & Selection                                                                         &     1048576 elements        & \begin{tabular}[c]{@{}l@{}}  PL, IL, DF, BR, BW, FR       \end{tabular}     
   \\\hline

\end{tabular}
}
\end{table}


\subsection{Evaluation Metrics}
\label{subsec:eval_metric}
\namePaper~is used to transfer a trained model using \textit{few-shot learning}. We then analyze the accuracy of the newly-built target model to predict the performance and resource \ommm{usage} of all the other configurations in 
\ommm{the target environment}. 
We evaluate the accuracy of the 
{transferred} model in terms of the mean relative error ($\epsilon_i$) to indicate the proximity of the predicted value $y_{i}^\prime$ to the actual value $y_{i}$ across \textit{N} test samples. The mean relative error (MRE) is calculated with Equation~\ref{eq:2}.
\begin{align}
 MRE= \frac{1}{N} \displaystyle \sum_{i=1}^{N} \epsilon_i=\frac{1}{N} \displaystyle \sum_{i=1}^{N}\frac{|y_{i}^\prime -y_{i}|}{y_{i}}    
\label{eq:2}
\end{align}

\section{\gs{Results}}
\label{sec:result}
\subsection{Accuracy Analysis \ommm{of the} Transferred Model}
\noindent {\textbf{Performance model transfer.}}
Figure~\ref{fig:acros_board} shows \namePaper's accuracy for transferring \omu{from edge to {cloud} platforms}. We make the following three observations. First, as we increase the number of labeled samples, the target model accuracy increases. However, the accuracy saturates and  with 5-10 \gs{samples} or \textit{shots}, we can achieve an accuracy as high as 80-90\%. Second, compared to applications with multiple complex kernels (\textit{blstm}, \textit{cedd}, \textit{digit} \omu{in Figure~\ref{fig:acros_board}(a), \ref{fig:acros_board}(b), and \ref{fig:acros_board}(c), respectively)}, simpler kernels (\textit{hist}, \textit{sc}, \textit{select} \omu{in Figure~\ref{fig:acros_board}(d), \ref{fig:acros_board}(e), and \ref{fig:acros_board}(f), respectively)} can be {more} easily transferred using fewer samples. \gs{There are two reasons for this trend: (1)} applications with multiple kernels have a larger optimization space. The large optimization space leads to more complex interactions that have compounding effects with other optimization options because we  
\ommm{model} multiple kernels rather than just a single~kernel, and
(2) simple kernels such as \textit{sc} and \textit{select} have been implemented using \textit{hls stream} interfaces, where rather than storing intermediate data in local FPGA memories, we read streams of data, and hence certain complex optimizations (like array partitioning) cannot be applied. This leads to a change in the feature space for different environments.
Third, \ommm{similar environments} are more amenable to transfer as the source, and target models are more closely related, \omu{i.e., we require small number of samples from the target environment to transfer a model. For example,} 
{transferring} to an FPGA with CAPI1 interface {(PCIe Gen3 with $\sim$~3.3 GB/s bandwidth)} 
from low-end PYNQ with PCIe Gen2 $\sim$~1.2 GB/s bandwidth entails a smaller \ommm{increase} in bandwidth 
than moving to an FPGA with CAPI2 interface {(PCIe Gen4 with $\sim$~12.3 GB/s bandwidth)}. 

Figure~\ref{fig:across_app} shows \namePaper's accuracy for transferring ML models \emph{across different applications}. \gs{We make three observations.}
First, as we increase the number of training samples, the target model accuracy increases for most of the applications. \gs{This observation is inline with the observation we made while transferring models performance models across different cloud platforms (Figure~\ref{fig:acros_board}).}
Second, the \ommm{largest}  improvement 
{in accuracy occurs} when our sample size $(c_{tl})$ 
{is} between 2 to 10. In most cases, {the accuracy} saturates after 20 samples. 
{Third}, in some cases, we see a decrease in accuracy {when increasing the number of samples}, e.g., Figure~\ref{fig:across_app}(a) for \texttt{HIST}, \texttt{SC}, and \texttt{SELECT}. This result could be attributed to training with a small amount of data, which can sometimes lead to overfitting~\cite{dai2007boosting}. 
\gs{We conclude that \namePaper is \ommm{effective at} transferring models \omu{from edge to} cloud platforms and \ommm{across} applications.}
\\

\begin{figure}[h]
\begin{subfigure}[b]{\linewidth}
\centering
   \includegraphics[width=\linewidth]{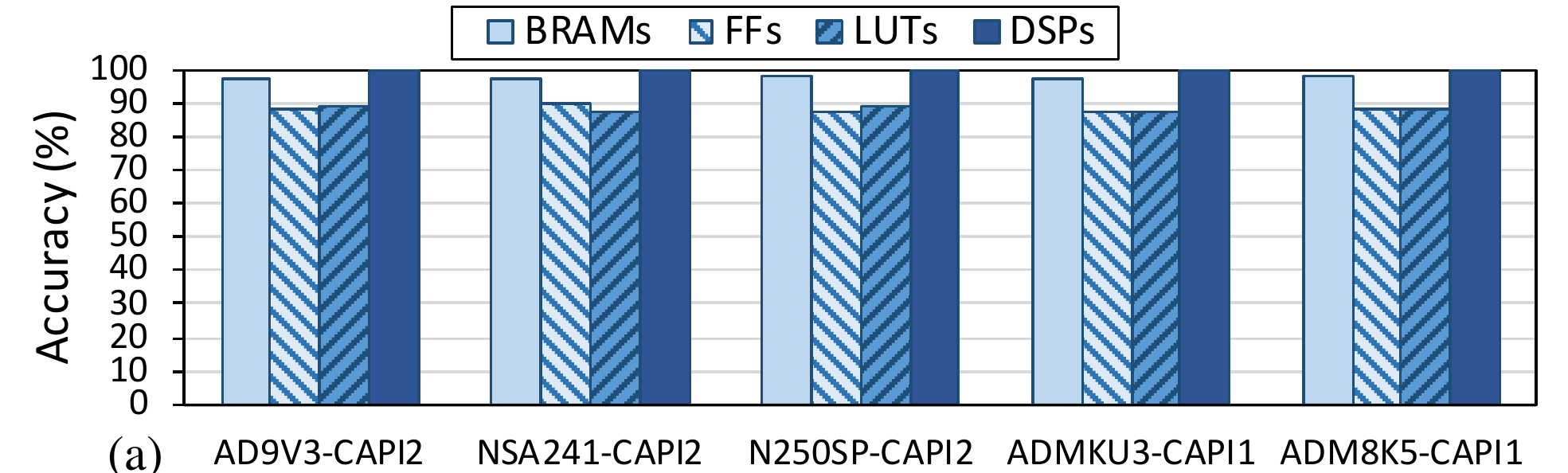}
         \phantomcaption
    \label{fig:area_board} 
\end{subfigure}
\begin{subfigure}[b]{\linewidth}
\centering
   \includegraphics[width=\linewidth]{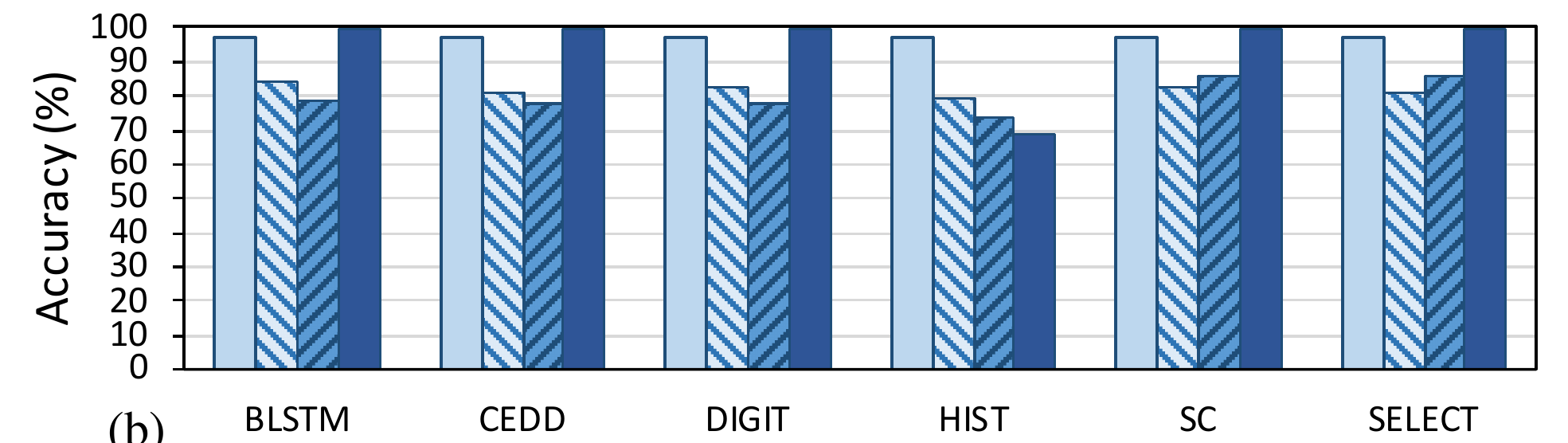}
      \phantomcaption
    \label{fig:area_app}

\end{subfigure}
\vspace{-0.9cm}
\caption[Two numerical solutions]{\namePaper's accuracy for transferring FPGA resource usage models through using (a) a base \ommm{model} trained on a low-end PYNQ-Z1 to different high-end target FPGA boards (horizontal axis), and  (b) different applications as base \ommm{models} (horizontal axis) to all the target applications, on low-end PYNQ-Z1 board.
}
\label{fig:area_transfer} 
\end{figure}

\noindent \textbf{Resource \ommm{usage} model transfer.} By using \namePaper, 
we can also train a resource \ommm{usage} model on a \textit{low-end} edge environment and transfer it to a high-end cloud environment.  Figure~\ref{fig:area_transfer} shows the accuracy of a target model trained by \textit{5-shot} {transfer} learning for predicting 
{a} resource \ommm{usage} vector $\eta_{\{BRAM, FF, LUT, DSP\}}$. The reported accuracy is for the transferred model, i.e., using a base model (low-end FPGA) to predict a target model (high-end cloud FPGA) after \gs{transfer} learning. In Figure~\ref{fig:area_transfer}(a), the horizontal axis depicts the target platform, while the base model is trained on a PYNQ-Z1 board. In the case of \ommm{across application} transfer (Figure~\ref{fig:area_transfer}(b)), the platform remains unchanged (PYNQ-Z1),  \ommm{while we use different applications to build the base model (horizontal axis).}

We make 
three observations. First, the resource \ommm{usage} model shows low error rates for predicting BRAM and DSP \ommm{usage}. This is attributed to the fact that the technological configuration of these resources remains relatively unchanged across platforms (e.g., BRAM is implemented as 18 Kbits in both the source and target platforms). Second, flip-flops and look-up-tables have 
comparably \ommm{lower accuracy} 
because the configuration of CLB slices varies with 
{the} transistor technology and FPGA family. {Third, while transferring across different applications (Figure~\ref{fig:area_transfer}(b)), we observe} relatively low accuracy for DSP \ommm{usage} while transferring a base model trained on \textit{hist}. This low accuracy is because the \textit{hist} accelerator implementation do\ommm{es} not use DSP units for logic. However, all other applications \ommm{use} DSP \ommm{units}. Therefore, the ML model trained on \textit{hist} \ommm{provides lower accuracy} in other environments \ommm{that use DSP units for accelerator implementation}. \gs{We conclude that \namePaper can efficiently transfer resource \ommm{usage} models.} 

\vspace{0.2cm}

In Table~\ref{tab:low_high_end}, we \ommm{report} the performance and resource \ommm{usage} for our \ommm{six} applications both on a low-end system and a high-end ADMKU3 FPGA-based cloud system. We use \namePaper to obtain the performance and resource \ommm{usage} for the high-end cloud configuration.
\begin{table}[h]
  \caption{{Execution time and resource \ommm{usage} for low-end edge configuration (PYNQ-Z1) and high-end cloud configuration (Nimbix \textit{np8f1} instance).}}
  \vspace{-0.4cm}
 \label{tab:low_high_end}
 \begin{center}
\renewcommand{\arraystretch}{0.85}
  \resizebox{\linewidth}{!}{%
\begin{tabular}{l|l|l|l|l|l|l}
\hline
\textbf{Application} & \textbf{Config.} & \textbf{Exec (msec)} & \textbf{BRAM} & \textbf{DSP} & \textbf{FF} & \textbf{LUT} \\ \hline
\multirow{ 2}{*}{blstm}         &      low-end              &   4200              &    80\%       &    15\%          &     24\%           &   47\%           \\
                                &\ommm{\namePaper}  &   1245              &    62\%       &    8\%          &     12\%           &   21\%           \\

\multirow{ 2}{*}{cedd}          &    low-end                &    10254              &     83\%          &    37\%          &   95\%          &     97\%         \\
                                &   \ommm{\namePaper} &    2217             &   56\%                &      3\%          &   75\%            &              94\%         \\

\multirow{ 2}{*}{digit}         &    low-end                &   2458             &     94\%              &       33\%         &     79\%          &            85\%     \\
                                & \ommm{\namePaper}  &       873          &     84\%              &       12\%         &     24\%          &            75\%              \\

\multirow{ 2}{*}{hist}          &    low-end                &    6173              &      94\%         &     0\%         &     11\%       &         37\%     \\
                                &  \ommm{\namePaper}  &   1104              &     67\%             &    0\%           &    5\%          &             30\%           \\

\multirow{ 2}{*}{sc}            &   low-end                 &   19306              &    82\%       &    0.4\%          &     12\%           &   25\%           \\
                                &\ommm{\namePaper}  &   4018              &    91\%       &    0.1\%          &     12\%           &   23\%     \\
 
\multirow{ 2}{*}{select}        &       low-end              &   18306              &    82\%       &    0.4\%          &     12\%           &   25\%           \\
                                &\ommm{\namePaper}  &   3918              &    91\%       &    0.1\%          &     12\%           &   23\%     \\   \hline
\end{tabular}
}
 \end{center}
\end{table}

\subsection{Base Model Accuracy Analysis}
\begin{figure}[h]
  \centering
  \includegraphics[width=\linewidth,trim={0.2cm 0.1cm 0.1cm 0.1cm},clip]{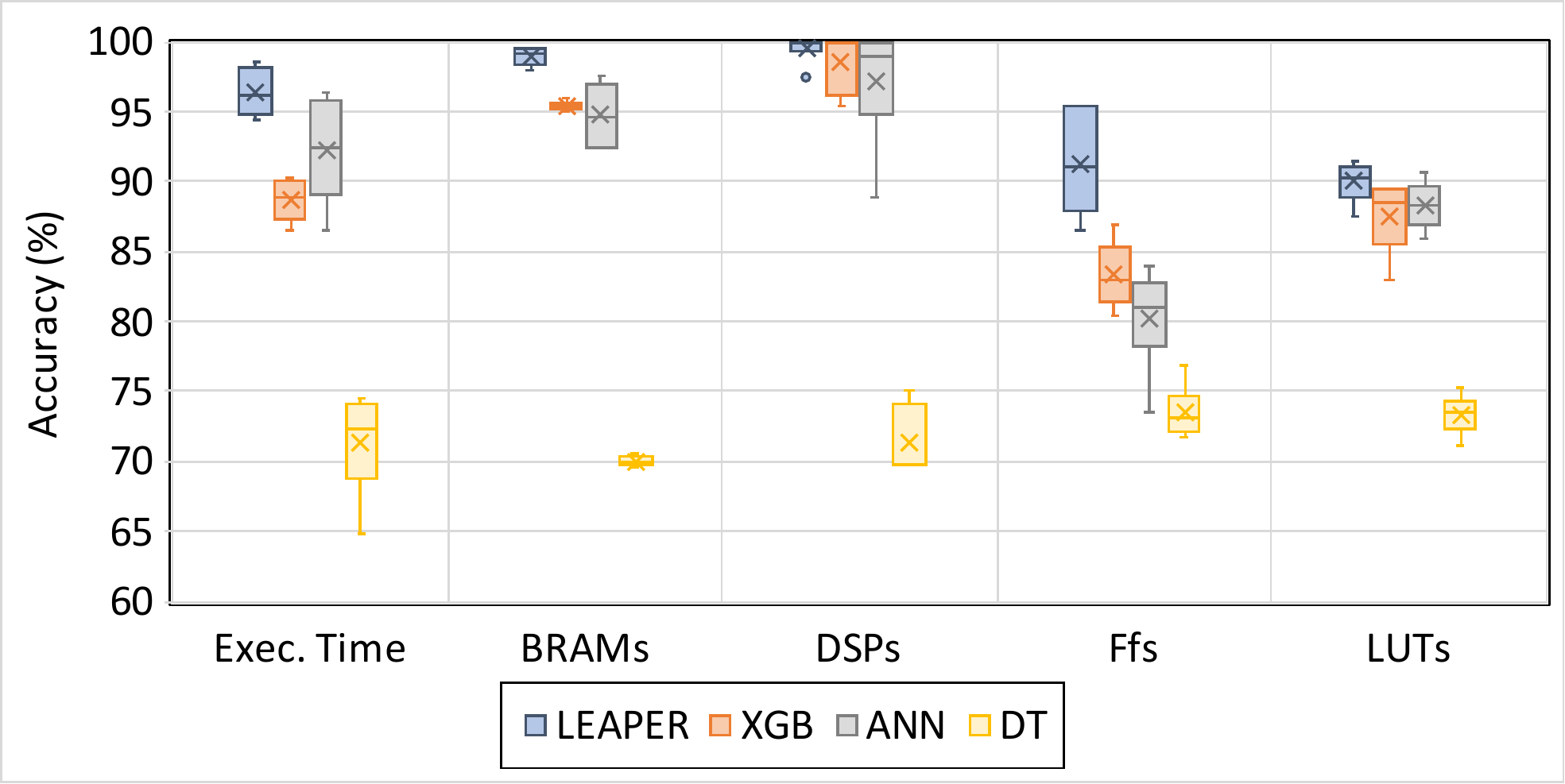}
  \vspace{-14pt}
  \caption{\ommm{Average} accuracy for performance (\texttt{Exec.}  \texttt{Time}) and resource  (\texttt{BRAM}, \texttt{DSP}, \texttt{FF}, \texttt{LUT}) \ommm{usage} predictions using \namePaper's base model and other machine learning techniques.}
  \label{fig:base_mode}
 \end{figure}

We also evaluate the accuracy of our base model. The base model is trained on our low-end edge PYNQ board using $c_{lhs}$ configurations sampled using \ommm{the} DoE technique. The base model can predict performance (or resource \ommm{usage}) outside the base model dataset (i.e, any configuration that is not a part of the DoE configuration space) $c_{lhs}$. To assess our base model, we use 30 previously unseen configurations that are not part of $c_{lhs}$ on the base system, and we evaluate the mean relative error for all 30 unseen configurations on all six applications.  Figure~\ref{fig:base_mode} shows the base model accuracy results. We also compare our base model to three other ML algorithms
that are also trained using $c_{lhs}$ configurations to predict performance and resource \ommm{usage}: XG-Boost (\texttt{XGB})~\cite{chen2016xgboost} based on  Dai~\etal~\cite{dai2018fast},
an artificial neural network (\texttt{ANN}) used by Makrani~\etal~\cite{xppe_asp_dac} and a traditional
decision tree (\texttt{DT})~\cite{loh2011classification}. 

 \vspace{0.2cm}
We make two major observations. First, on average, \namePaper~is \gss{8.1\% (4.1\%),	4.3\% (5.1\%), and	25.9\% (23.9\%) more accurate in terms of performance (resource \ommm{usage}) prediction than \texttt{XGB}, \texttt{ANN}, and \texttt{DT}, respectively. 
Second, \texttt{ANN} is  22.7\% and 19.5\% more accurate than  \textit{DT} for performance and resource \ommm{usage} prediction, respectively, but performs \ommm{worse}
than \namePaper. \texttt{ANN} is not sample-efficient as it requires more training samples to learn~\cite{li2018processor}.}  \gs{We conclude that \namePaper's base model provides both high accuracy and sample-efficiency compared to other ML-based algorithms.}

We also compare the performance of using different \ommm{ML-based methods} for transferring models across different platforms and across different applications.
In Table~\ref{tab:compare}, we show the average accuracy of \namePaper's ensemble of transfer learner and compare it to two different \ommm{ML-based methods}: (1) decision tree (\texttt{DT})~\cite{loh2011classification}, and (2) adaBoost (\texttt{ADA})~\cite{adaboost}. 
We observe that \namePaper is \ommm{on average} 12.1\% (10.6\%) and 6.6\% (7.7\%)  on average more accurate in transferring models across platforms (across applications) than \texttt{DT} and \texttt{ADA}, respectively. We conclude that an ensemble of \ommm{transfer learner} is better than using a single \ommm{transfer learner} while transferring across different FPGA-based environments.
 
 \vspace{0.2cm}
\begin{table}[h]
\centering
  \caption{Average accuracy (\%) comparison of \namePaper~ with decision tree (DT) and adaBoost (ADA) as TL for~\textit{5-shot}~transfer.}
    \vspace{-0.1cm}
    \label{tab:compare}
    \resizebox{0.6\linewidth}{!}{%
\begin{tabular}{@{}llll@{}}
\hline
\textbf{Environment}               & \textbf{DT}   & \textbf{ADA}   & \textbf{\namePaper} \\ \hline
Across Platform       & 77.7 & 83.2 & 89.8  \\
Across Application & 70.6 & 73.5 & 81.2   \\ \hline
\end{tabular}
}
\end{table}

\subsection{Target {Cloud FPGA }Model Building Cost}
\label{subsection:timing}
\gs{In Table~\ref{tab:timing}, we mention the time and cost to build a model from scratch on a cloud environment using the traditional \ommm{ML-based} approach and compare it to using \namePaper to build a model for the cloud environment.}
If we build a model from scratch, then we need 50 sampled DoE configurations ($c_{lhs}$) for which the time and cost is mentioned in \textit{DoE run 
(hours)} and \textit{DoE cost}, respectively. Table~\ref{tab:timing} also includes the {execution} time 
on the ADMKU3 {cloud} platform (``Exec (msec)''. While the process of synthesis and place and route (P\&R) for the cloud FPGA, which is needed to obtain performance estimates in terms of maximum operating clock frequency and the resource \ommm{usage}, can be carried out offline, most of the cloud providers offer virtual machines (VMs) with all the appropriate software, IPs, and licenses needed to generate an FPGA image ready to be deployed at their cloud infrastructure (e.g., the Vivado AMI of AWS~\cite{AMI}). Therefore, we include the cost of the cloud environment (``Est. Cost (\$)'') for data collection.

By using \namePaper, the DoE runtime is amortized and, by using a few labeled samples $c_{tl}$ (``5-shot (hours)'') from the target platform, we can transfer a previously trained model and make predictions for all the other configurations for the target platform. We mention the the transfer time (``Transfer (msec)'') for each model. 
{As a result}, quick exploration and {significant} time savings {(at least 10.2$\times$) are possible when transferring a model (i.e., ``5-shot (hours)'' + ``Transfer (msec)'') as compared to building a new model from scratch (i.e., ``DoE run (hours)'') + ``Exec (msec)'')}. DoE reduces training samples from 500+ to 50, while  \textit{5-shot} transfer learning further reduces the number of samples to 5, so we achieve $\sim100\times$ effective speedup compared to a traditional ``brute-force'' approach for data collection.

\vspace{0.3cm}
\begin{table}[h]
\caption{DoE time for gathering sampled data points for a single CPU-FPGA platform (``DoE run (hours)''), DoE execution time on the deployed platform (``Exec (ms)''), Estimated Cost on a cloud platform (``Est. Cost (\$)''), time for gathering 5 labeled samples (``5-shot (hours)''), \namePaper~time including the transfer time (``Transfer (msec)''), ``Speedup'' over building a new model from scratch using only the DoE data
.}
\label{tab:timing}
\vspace{-0.1cm}
\resizebox{1\linewidth}{!}{%
\begin{tabular}{p{0.1\linewidth}|lll|lll|l}
\hline
\multicolumn{1}{c|}{\textbf{Application}}  & \multicolumn{3}{c}{\textbf{Traditional}} &
\multicolumn{3}{c}{\textbf{\namePaper}} &
\\ 

Name &
 \begin{tabular}[c]{@{}l@{}}       
 DoE run \\(hours)

\end{tabular}&
 \begin{tabular}[c]{@{}l@{}}       
Exec \\(msec)

\end{tabular}&
 \begin{tabular}[c]{@{}l@{}}       
Est.Cost$^{7}$\\(\$)

\end{tabular}&
 \begin{tabular}[c]{@{}l@{}}       
5-shot \\(hours)

\end{tabular}
& \begin{tabular}[c]{@{}l@{}}       
Transfer \\(msec)

\end{tabular}&
 \begin{tabular}[c]{@{}l@{}}       
Est.Cost$^{7}$\\(\$)

\end{tabular}
&Speedup\\ \hline

blstm   &  135  &  455 & 168.7 & 13 & 55.6 & 16.2 & 10.4  \\

cedd    &  124  & 295 & 155.0 & 12 & 26.5 & 15.0 & 10.3 \\

digit   & 122   & 435 & 152.5 & 12 & 58.8 & 14.9 & 10.2\\

hist    & 97  &  45  &  121.2 & 9 & 17.1   & 11.3 & 10.8   \\

sc      & 104  & 145 &  130.0 & 10 & 27.9 & 12.4 & 10.4 \\

select  &  106   & 145 & 132.5 & 10 & 27.6 &  12.5 & 10.6
\\

   \hline 
   \multicolumn{8}{l}{$^{7}$\small{The cost is estimated based on an enterprise online cost estimator~\cite{nimbix-calc}. Specifically, we selected an}} \\
   \multicolumn{8}{l}{\small{\textit{n2}~(8-core, 64GB RAM VM - 1.25\$/h) for bitstream generation (x86) and an \textit{np8f1} instance (160-thread}} \\
   \multicolumn{8}{l}{\small{1TB RAM, ADM-PCIE-KU3 with CAPI-1 - 3\$/h) for deployment.}} \\
\end{tabular}
}
\end{table}







\section{\gs{Explainability: Why does \namePaper~work?}}
 \label{section:relatedness_analysis}
{To explain our results for transfer learning}, we analyze the degree of \textit{relatedness} between the source and target environments. 
We use two different analysis techniques: (1) divergence analysis~\cite{coutinho2011divergence}, and (2) correlation analysis~\cite{benesty2009pearson}.

\head{Divergence analysis} We use Jensen-Shannon divergence (JSD)~\cite{jsd} to measure  the difference between two probability distributions of the source ($P(\tau_{s})$) and the target environment ($P(\tau_{t})$). 
The lower the JSD \ommm{value}, the more similar the target environment is to the source (i.e., if $D_{JSD}(P(\tau_{t}) || P(\tau_{s})=0$ implies that the distributions are identical and 1 indicates unrelated distributions). Table~\ref{tab:jsd} shows the JSD analysis for transferring models between different applications. We make three main observations. First, JSD analysis confirms the trend observed from transferring application models (Figure~\ref{fig:across_app}), i.e., the more closely related the source and target applications, 
the fewer 
samples are required to train our non-linear transfer learners. Second, the higher the JSD between two applications, the lower the accuracy while transferring between those tasks. 
Third, for many applications JSD values is low, which indicates that we can easily transfer models between such environments using a few samples from the target environment. 

 \vspace{0.2cm}
\begin{table}[h]
\centering
  \caption{Jensen-Shannon Divergence (JSD)~\cite{jsd} between 
  performance distributions of different applications. JSD measures statistical distance between two probability distributions.}
  \vspace{-0.1cm}
    \label{tab:jsd}
     \renewcommand{\arraystretch}{0.9}
\resizebox{0.75\linewidth}{!}{%
\begin{tabular}{@{}clllllll@{}}
\hline   
\multirow{10}{*}{\rotatebox{90}{\textbf{Target Model}}} & \multicolumn{7}{c}{\textbf{Base Learner}}\\

                              & \multicolumn{1}{l}{}       & blstm & cedd & digit & hist & sc   & select \\ 
                              \cmidrule{2-8} 
                              & \multicolumn{1}{l|}{blstm}  & 0.00     & 0.24 & 0.34  & 0.25 & 0.31 & 0.30   \\
                              & \multicolumn{1}{l|}{cedd}   & 0.24  & 0.00     & 0.49  & 0.54 & 0.41 & 0.40   \\
                              & \multicolumn{1}{l|}{digit}  & 0.34  & 0.49 & 0.00      & 0.25 & 0.21 & 0.21   \\
                              & \multicolumn{1}{l|}{hist}   & 0.25  & 0.54 & 0.25  & 0.00     & 0.25 & 0.24   \\
                              & \multicolumn{1}{l|}{sc}     & 0.30  & 0.40 & 0.21  & 0.24 & 0.00     & 0.05   \\
                              & \multicolumn{1}{l|}{select} & 0.30  & 0.41 & 0.21  & 0.25 & 0.05 & 0.00       \\ \hline
\end{tabular}
}
\end{table}

\head{Correlation analysis} \ommm{Correlation analysis} measures the strength of linear correlation between two environments. We make four major observations. First, for different target hardware platforms, we have a high correlation of $0.76$ to $0.97$ {between the source and target  execution time}, which indicates that the target model's performance behavior can \ommm{accurately} be learned using the source environment.  
Second, as we switch to a higher external bandwidth {for the target platform (i.e., CAPI1 to CAPI2)}, 
the correlation becomes lower because the hardware change is much more \textit{severe} coming from a low-end FPGA with limited external bandwidth.  \omu{Whereas changing the technology node from one FPGA to another (e.g., changing from ADMKU3 board to  AD9V3 board) leads to a smaller change in the environment because of the linear relation between technology node and performance. }
Third, the correlation between applications on a single platform is lower ($0.45$ to $0.9$) because of {the} varying application characteristics {and optimization space.} Fourth, as the linear correlation is not 1 for all platforms, the use of a nonlinear transfer models is substantiated. We conclude that \namePaper learns differences in environments to accurately transfer FPGA-based system  performance prediction models \omu{from one platform to another.}




\section{Discussion and Limitations}
 \label{section:limit}

\head{\namePaper's~generality}
\namePaper is a framework for building and transferring models from a small edge environment to any new, unknown FPGA-based environment. We demonstrate our approach using the cloud system as our target environment because cloud systems often use expensive, high-end FPGAs, e.g., Amazon AWS~F1 cloud~\cite{aws}, Alibaba Elastic cloud~\cite{alibaba}, etc. We can, thus, achieve tangible gains in terms of cost, efficiency, and performance. However, \namePaper can be used to transfer models to any high-end FPGA system.

\noindent\textbf{Effect of FPGA resource saturation.}
An FPGA gives us the flexibility to map \ommm{a given} operation to different \ommm{potential} resources. For example, we can map a multiplication operation  to either a CLB or a DSP slice. \ommm{We can decide the mapping based on the operand width, i.e., if the operand width is smaller than DSP slice width, the operation is mapped to a CLB otherwise to a DSP unit.} 
\ommm{Currently, we do not consider mapping the same operation to different resources.}


\head{Transfer a model to a new platform and application simultaneously}
In supervised learning, transferring both to a new platform and application at the same time would lead to sub-optimal results (as observed in~\cite{xapp}). This sub-optimal performance is because in such a scenario we would perform two \ommm{types} of transfer \ommm{at the same time to (1)} unknown hardware and (2) unknown application. 
We explicitly exclude this scenario in the current work.
\section{Related Work}
To our knowledge,  {\namePaper} is the first work to {leverage an ML-based performance and resource {usage} model trained for a low-end \om{edge environment} to predict \ommm{the} performance and resource {usage} of an accelerator implementation for a new, high-end cloud environment.}  FPGAs  {lead to} very low productivity due to the time-consuming downstream accelerator implementation process. 
In this section, we describe other related works in ML-based modeling \ommm{of FPGA}, analytical  modeling \ommm{of FPGA}, and transfer learning.

\noindent\textbf{ML-based modeling  \ommm{of FPGAs}.} Recent works propose ML-based methods~\cite{makrani2019pyramid,ferianc2020improving,ustun2020accurate,o2018hlspredict,xppe_asp_dac,dai2018fast,zhao2019machine,yanghua2016improving,wang2020machine,mahapatra2014machine,sun2022correlated} to overcome the issue of low productivity while designing FPGA-based accelerators. O'Neal~\etal~\cite{o2018hlspredict} use CPU performance counters to train several ML-based models to predict \omu{the} performance and power consumption of an accelerator implementation. Makrani~\etal~\cite{xppe_asp_dac} train a neural network-based model to predict application speedup across different FPGAs. Makrani~\etal~\cite{makrani2019pyramid} and Dai~\etal~\cite{dai2018fast} use ML to predict resource \ommm{usage} for an accelerator implementation. 
However, these solutions become largely impractical once the platform, the application, or even the size of the workload changes. \namePaper~proposes to reuse previously\ommm{-}built models for a low-end source environment on a high-end target environment through transfer learning. Unlike \namePaper, past works apply traditional, time-consuming brute-force techniques to collect\omu{s} training datasets. 

\noindent\textbf{Analytical modeling  \ommm{of FPGAs}.} {Analytic\ommm{al} models abstract low-level system details and
provide quick performance estimates at the cost of accuracy. These approaches \ommm{(e.g.,}~\cite{linAnalyzer,zhao2017comba,hlsscope,chai_icpe19}) analyze dataflow graphs and apply mathematical equations to approximate resource usage or performance after the HLS pre-implementation phase. \ommm{Even though they} enable {quick early-stage design studies, however, analytical models} are not able to model the intricacies of the complete  implementation process~\cite{dai2018fast}. Therefore, these approaches provide crude estimates of the actual performance. Moreover, these models require \ommm{FPGA domain} knowledge to form mathematical equations.  In contrast,~\namePaper~does not require expert knowledge to construct equations. \namePaper~learns from the \ommm{training} data \ommm{(application features and accelerator optimization options) to consider} the complete downstream accelerator implementation process and provides the capability to transfer models from an edge-FPGA to a high-end cloud FPGA environment.}

\noindent\textbf{Transfer learning.} Recently, transfer learning~\cite{chen2010experience,pan2009survey,baumann2018classifying,bozinovski1976influence,pratt1992discriminability} has gained traction to decrease the cost of learning by transferring knowledge \ommm{between different tasks}. 
Valov~\etal~\cite{valov2017transferring} investigate the transfer of application models across different CPU-based environments using linear transformations. Jamshidi~\etal~\cite{jamshidi2017transfer} demonstrate the applicability of using nonlinear models to transfer CPU-based  performance models. {The works above influenced the design of~\namePaper}. In contrast \ommm{to them}, we:  (1) focus on FPGA-based systems that, unlike a CPU-based system, ha\ommm{ve} a different hardware architecture for every application and optimization strategy, and 
(2) use an ensemble of transfer learners that transfers accurate models to a target environment. \\
\section{Conclusion}

We introduce \namePaper, {the first} \textit{transfer learning}-based approach \ommm{for prediction of performance and resource usage in FPGA-based systems. 
\namePaper overcomes the inefficiency of traditional ML-based methods by leveraging an ML-based performance and resource \ommm{usage} model trained for a low-end \om{edge environment} to predict \omu{the} performance and resource \ommm{usage} of an accelerator implementation for a new, high-end cloud environment.}

 
 Our experiments show that \ommm{\namePaper is} cheaper (with \textit{5-shot}), faster (up to $10\times$), and highly accurate (on average 85\%)  \ommm{at predicting} performance and resource \ommm{usage} in a new, {unknown} target {cloud} environment \ommm{than building models from scratch}. 
 We believe that \namePaper~\ommm{can} open up new avenues for research on FPGA-based systems from edge to cloud computing, and hopefully, \ommm{it will} inspire the development of new modeling techniques \ommm{for FPGAs}. 










\section*{Acknowledgments} 
\label{sec:acknowledgment}
 \ommm{We thank the SAFARI Research Group members for valuable feedback and the stimulating intellectual environment they provide. Special thanks to Florian Auernhammer and Raphael Polig for providing access to IBM systems. We appreciate valuable discussions with Kaan Kara.} This work was performed in the framework of the Horizon 2020 program for the project ``Near-Memory Computing (NeMeCo)''. It is funded by the European Commission under Marie Sklodowska-Curie Innovative Training Networks European Industrial Doctorate (Project ID: 676240). We acknowledge the generous gifts of our industrial partners, especially  Google, Huawei, Intel, Microsoft, VMware. This research was partially supported by the Semiconductor Research Corporation and the ETH Future Computing Laboratory.
\balance
\bibliographystyle{IEEEtranS}
{\small\bibliography{ref}}

\end{document}